\documentclass[manuscript]{aastex}

\slugcomment{}

\shorttitle{Emission-line Star Survey of the IC 1396 H {\sc ii} region}
\shortauthors{Nakano et al.}

\begin{document}


\title{Wide-Field Survey of Emission-line Stars in IC 1396}


\author{M. Nakano}
\affil{Faculty of Education and Welfare Science, Oita University,
Oita 870-1192, Japan}
\email{mnakano@oita-u.ac.jp}

\author{K. Sugitani}
\affil{Graduate School of Natural Sciences, Nagoya City University,
Mizuho-ku, Nagoya 467-8501, Japan}

\author{M. Watanabe}
\affil{Department of Cosmosciences, Hokkaido University,
Sapporo 060-0810, Japan}

\author{N. Fukuda}
\affil{Department of Computer Simulation, Okayama University of Science, 
1-1 Ridai-cho, Okayama 700-0005, Japan}

\author{D. Ishihara}
\affil{Department of Physics, Nagoya University, Furo-cho, Chikusa-ku, Nagoya 
464-8602, Japan}

\and

\author{M. Ueno}
\affil{Institute of Space and Astronautical Science, Japan Aerospace Exploration Agency, 
3-1-1 Yoshino-dai, Chuo-ku, Sagamihara 252-5210,  Japan}




\begin{abstract}
We have made an extensive survey of emission-line stars in the IC 1396 H {\sc ii} region to  investigate the 
low-mass population of pre-main sequence (PMS) stars.
A total of 639 H$\alpha$ emission-line stars were detected in an area of  4.2 deg$^2$ 
and their  $i'$-photometry was measured.
Their spatial distribution exhibits several aggregates near the elephant trunk globule (Rim A)
and bright-rimmed clouds at the edge of the H {\sc ii} region (Rim B and SFO 37, 38, 39, 41), 
and near HD 206267, which is the main exciting star of the H {\sc ii} region.
Based on the extinction estimated from the near-infrared (NIR) color--color diagram, we have selected pre-main sequence 
star candidates associated with IC 1396. The age and mass were derived from the extinction corrected color-magnitude 
diagram and theoretical pre-main sequence tracks.
Most of our PMS candidates have ages of $<$ 3 Myr and masses of 0.2--0.6 M$_\sun$. 
Although it appears that only a few stars were formed in the last 1 Myr in the east region of the exciting star,
the age difference among subregions in our surveyed area is not clear from the statistical test.
Our results may suggest that massive stars were born after the continuous formation
of  low-mass stars for 10 Myr.
The birth of the exciting star could be the late stage of 
slow but contiguous star formation in the natal molecular cloud.
It may have triggered to form many low-mass stars at the 
dense inhomogeneity in and around  the H {\sc ii} region  by a radiation-driven implosion.

\end{abstract}





\section{Introduction}

As observational instruments have developed,  systematic studies of
stellar clusters have revealed  
many low-mass populations of pre-main sequence (PMS) stars in massive star-forming regions.
It has been found that most stars are born in groups or clusters, not in isolated regions \citep[e.g.,][]{lad03, all07}.   
Stellar clusters
provide a tool for investigating stellar evolution, because
they are considered to constitute a natural sample of coeval stars at the same distance 
and with the same chemical composition.
Young clusters appear to be  a snapshot of stellar evolution over a wide range of masses. 
However, 
a careful study of the Orion Nebula Cluster \citep{hil97}  has revealed 
that the younger stars ( $<$ 0.3 Myr )  are clustered towards the center of the Trapezium Cluster
(a projected distance of $<$ 0.3 pc from ${\theta}^1$C Ori), while the older stars are widely distributed.  
This indicates that star formation has recently occurred in the center of the cluster and that star formation
with a low rate has taken place over a long time scale over the whole region.
\citet{mue08} has suggested that there is no physical reason to
separate the Orion Nebula Cluster and the Trapezium Cluster, and the entire
region is a single contiguous star forming event. 
There are also examples of stellar clusters still producing massive stars.
Cep A HW2 is the brightest radio source in Cep A \citep{huw84}, 
an active star forming region within the molecular cloud associated with the cluster 
with an age of 5.5 Myr \citep[Cep OB3b;][]{jor96}.  
\citet{Jim07} found a hot core internally heated by a massive protostar, 
and a circumstellar rotating disk around HW2.  
This suggests that a massive star is still forming in the young cluster.
Furthermore, X-ray stars with a mid-infrared (MIR)  spectral energy distribution of 
young stellar objects (YSOs) around HW2 have been reported by \citet{prv09}.
It is most likely that a low-mass PMS population was already formed before the formation of high-mass stars.
Once a high-mass star starts to radiate significant amounts of UV radiation,  
the surrounding material is affected dramatically and the natal molecular cloud is disrupted
and the birth of the next generation of stars might be triggered.

The apparent age spread observed in the color-magnitude diagram of young stellar 
groups has been investigated by \cite{bur05}. 
They found that the spread could not be explained adequately by the combination of the binarity, 
photometric uncertainty and variability of the members. 
In the case of the $\sigma$ Ori young stellar group, 
they suggested an age spread of 2--4 Myr.
A possible explanation is that accretion from a disk causes 
scatter of the stellar continuum light \citep{tou99}.
However, another possibility is that  stars in a cluster are 
formed over a period of time.  
Three-dimensional numerical simulations including protostellar outflow feedback   
show that the cluster-forming clumps can maintain a quasi-virial equilibrium state 
for several periods of free-fall time \citep{nal07}, suggesting continuous star formation.  
\citet{wan10}  also demonstrated  outflow-regulated massive star formation.
The episodic star formation causes an intrinsic spread in age and hence in the position in the color-magnitude diagram. 

We have reported that young stars near the exciting star of the W5E H {\sc ii} region are
systematically older than those near the edge of the H {\sc ii} region  \citep{nak08}.
We further suggested that the birth of the most massive star occurs in the late stage of star formation in the cluster.
Although the structure of the H {\sc ii} region is simple to investigate, the distance of W5 is large, at 2.3 kpc,
and therefore it is not easily accessible
to the wide range of stellar masses.

IC 1396 is a ring-shaped giant H {\sc ii} region in Cep OB2 with a size of about 3$\degr$, 
located at the edge of the Cepheus bubble, which is a large, far-infrared shell with a diameter
of $\sim$10$\degr$ \citep{kun08}. 
The structure of  IC 1396 is rather simple and is powered by the 
multiple O star system HD 206267 (RA = 21$^h$38$^m$57$\fs$62, Dec = +57$\arcdeg$29$\arcmin$20$\farcs$5), which is located near the 
geometrical center of IC 1396  in the young cluster Trumpler 37. 
The bright members of Trumpler 37 have been investigated by many authors \citep[e.g.,][]{gar76, mar90, dez99, con02}.
There are a number of bright-rimmed globules \citep{pot56} at its periphery.
Its modest distance ($<$ 1 kpc) and low foreground extinction ($A$v $<$ 1 mag) make IC 1396 a good subject for 
our study of low-mass population associated with OB stars.
The mean distance of 76 members of Cep OB2 derived from the Hipparcos parallax is 615 pc  \citep{dez99}.  
\citet{con02} made a color-magnitude diagram of the intermediate-mass members and fitted them to
main-sequence stars at 870 pc.   
In this paper, we adopt 870 pc, which is frequently quoted by the recent works  \citep[e.g.,][]{sic04} .

H$\alpha$ emission, a characteristic of T Tauri stars (TTS), is radiated from the solar-type 
magnetic activity in weak-lined T Tauri Stars (WTTS) or the chromospheric activity and magnetospheric accretion shock in 
classical T Tauri stars (CTTS)  \citep{bri07}. 
Thus, a survey of H$\alpha$ emission-line stars is a powerful method for searching 
for young low-mass populations \citep[e.g.,][]{nak95}.  
Recently, NIR observations with  array detectors and   
MIR observations from space (e.g. with  the {\it Spitzer} Space Telescope) have advanced
the protostar survey in the dense parts of giant molecular clouds.  
However, the H$\alpha$ survey is still useful in areas with moderate or low extinction, and it is also a complementary 
method to search for PMS populations in a wide and rather dispersed region.
Here, we report an extensive survey of emission-line stars in the IC 1396 H {\sc ii} region and discuss
the formation sequence of low and high mass populations in IC 1396.

\section{Observations}

Slitless grism spectroscopic searches for H$\alpha$ emission objects in IC 1396 were conducted 
in 2004 and between 2006 and 2009 using the Wide Field Grism 
Spectrograph 2  \citep[WFGS2; ][]{ueh04} at the f/10 Cassegrain focus of the 
University of Hawaii (UH) 2.2-m telescope on Mauna Kea.  
The log of the observations is given in Table 1.
A 300-line mm$^{-1}$ grism was used in conjunction with a wide H$\alpha$ filter to isolate 
the first-order spectra between 6265--6765 {\AA}.  
A Tektronix 2048$\times$2048 CCD was used as the detector, yielding a dispersion of  
3.8 {\AA} pixel$^{-1}$, a pixel scale of 0$\farcs$34  and an instrumental field of 
view of 11$\farcm$5 $\times$ 11$\farcm$5


The whole IC 1396 region was covered by 157 field positions including overlapping areas. 
About 4.2 deg$^2$ were observed, both in the grism and the direct imaging mode of WFGS2. 
Figure 1 shows the boundary of the observed fields and indicates the names of bright-rims,  dark clouds, 
and two O stars in the IC 1396 H {\sc ii} region.   
For each field, we have taken a single 30-s exposure in the direct mode with a wide H$\alpha$ filter.
Then, we inserted a grism and obtained  three dithered frames of 300-s exposures  for 
slitless spectroscopy. 
Direct images with an $i'$-band filter were also taken with WFGS2 in 2008 August and 2009 August. 
Three 30-s exposures were made for each field. 
The $i'$-band flux is less affected by  excess emission of the circumstellar disk, H$\alpha$ emission, and
continuum veiling of CTTS.
Thus,  $i'$-band photometry is useful as a measure of the underlying photosphere of CTTS
\citep[e.g.,][]{cie05}. 
3--5 standards for the Sloan Digital Sky Survey (SDSS) system \citep{smi02} were observed each 
night at nearly the same 
airmass (difference of $<$ 0.1) as the targets.   We used a dome flat for flat fielding. 

The data were reduced using the IRAF software package. 
We applied the standard procedure for bias subtraction and flat fielding.
For each field, three dithered grism frames are combined into one frame by referring the position of 
the H$\alpha$ emission or absorption line of the star(s) in the field.
We picked all stars in each wide H$\alpha$ image using the DAOFIND package 
and extracted their one-dimensional spectra from the grism
image. We visually investigated the spectra for the presence of the
H$\alpha$ emission line and found 639 H$\alpha$ emission-line objects in the whole surveyed area.  
We used the SPLOT package to measure the equivalent width 
of the H$\alpha$ line in the spectra.  Their coordinates were determined by the astrometric 
calibration tool of the Starlink program {\it GAIA} using  USNO-B  catalog stars as the reference in the field.
All of the photometric measurements were made with the APPHOT package for aperture photometry.  
The typical seeing of the image was 2.7--3.0 pixels. 
We used  a 4-pixel diameter aperture for the aperture photometry.
We then calibrated the photometric values to the SDSS system via standards. 
In general, the run was of high quality, but thin cloud affected a few fields.
We checked the stability of the magnitude for three direct images in each field and examined the quality of the measurement. 
Excluding stars with saturated images or images with low photometric quality, we measured 586 stars in 13.6  $<$ $i'$ $<$ 19.3.

\section{Results}

\subsection{Spatial Distribution}

We  list the 639 H$\alpha$ stars in Table 2. 
The table shows the (1) identification number, (2, 3) J2000 coordinates, (4) $i'$-band magnitudes and 
(5) equivalent widths of H$\alpha$ emission ({\AA}).
The H$\alpha$ equivalent widths were measured for 608 stars.
We compared our list of emission stars with the Two Micron All Sky Survey (2MASS) Point Source Catalog  
with 12 stars supplemented by the 2MASS 6x Point Source Catalog.  
We found matches for 617 stars with 2MASS stars within 2$\arcsec$ and determined that 548 of these have good quality $JHKs$ photometry (quality flag AAA). 
Due to small separation angles or positional uncertainties,
two pairs of  H$\alpha$ stars are identified with single 2MASS sources
(No. 250/251 with J21381703+5739265, and No. 600/601 with J21451578+5717392).   
We have included these stars in the above number. 
We list the 2MASS designation of the NIR counterpart, photometric quality, $J$ magnitude, and $J$--$H$, $H$--$Ks$ colors  
in columns (6)--(10) of Table 2.
$A$v estimated from the NIR color--color diagram (see \S 4.1) is also listed in column (11). 
About 70\% of our H$\alpha$ stars cannot be identified with known sources in 
the SIMBAD database. 
The identification in the literature or comments are listed in column (12).


Figure 2a shows the spatial distribution of emission-line objects
while Figure 2b represents the contour map of the number density of the detected emission-line objects with 
the kernel method \citep{gom93}.  We adopted a Gaussian shape for the density distribution 
of the kernel with a smoothing parameter of h$=$2$\farcm$4.
The isolines are drawn at intervals of 150 stars per deg$^2$, which is about three times the average number density
in the field region (see Table 4), starting from 100 stars per deg$^2$.  
It is clear from Figure 2 that the distribution of emission-line stars is far from uniform.  
To demonstrate 
the different properties in the different areas, we divided the surveyed field into 
the following subregions according to the number density of H$\alpha$ stars. 
The primary peak of the most rich aggregate (the West region) is located between the main exciting star 
HD 206267 and  the western elephant trunk globule Rim A (IC 1396A). 
The secondary peak is at the opposite side, to the east of HD 206267 (the East region),  and
the third peak is at the south of the bright-rimmed globule SFO 38  (SFO: \citealt{sfo91}, the SFO38 region).  
Other small aggregates of emission-line stars are located near SFO 37 (the SFO37 region), SFO 39 and SFO 41 
(the SFO39/41 region), Rim B (the Rim B region), and at the north of the B 365 dark cloud (B: \citealt{b27}) 
or around (21$^{h}$35$^{m}$, $+$57$\degr$).  See Figure 1 for references.  
The West and the East regions appear to be surrounded by a halo-like structure (the Halo region). 
Some H$\alpha$ stars appear to be associated with other small bright rimmed clouds 
such as Rim C and SFO 34 (Rim D).   
We defined the rest of the region as the field area (the Field region).

\subsection{Source Identification }

Our sample of H$\alpha$ stars  contains  two carbon stars CGCS 5454 (No. 575) and CGCS 5401 (No. 287),  
and one Be/X-ray binary Cep X-4 (No. 340), whose distance was confirmed to be 3.8 kpc by \citet{bon98}.
Our sample of H$\alpha$ stars also contains a candidate of the exciting source  of the Herbig--Haro object 
HH 588 (No. 436), which is deeply embedded in 
SFO 37 \citep{ike08}, one Herbig Ae/Be star (HAeBe) MVA 426 (No. 236) \citep{sic07},  and $\sim$ 100 TTS 
including Class-II, III, and transient objects  \citep{sic06b, sic07, get07, rea04, mer09, bar11}. 

An objective prism survey using the Schmidt telescope \citet{kun86} detected 155 H$\alpha$ emission objects in IC 1396
and an additional 65 emission stars were reported in a subsequent paper \citep{kup90}.
Their spatial distribution shows no noticeable concentration or clustering. 
Out of 220 Kun H$\alpha$ stars, 150 stars are located in our observed field; however, 
only 11 stars (No. 287 = Kun 80, No. 13 = Kun 171, No. 340 = Kun 193, No. 618 = Kun 215, 
No. 49 = Kun 308, No. 80 = Kun 310, No. 97 = Kun 314 S, No. 252 = Kun 321, No. 318 = Kun 325, 
No. 410 = Kun 330, No. 487 = Kun 332) are identified among our H$\alpha$ stars.  
Although Kun 315 (LkH$\alpha$ 349) is not included in Table 2, our grism spectrum appears saturated 
but indicates a P Cyg-type H$\alpha$ profile. 
Kun 320 (GL Cep) and Kun 337 (HD 239745) are also too bright 
and saturated in our grism spectra, and the positions of Nos. 77, 230, 12, and 266 are offset 
from Kun 53, 73, 170, and 322 by around 10$\arcsec$.  
Even if we include these seven stars, only 12\% 
of stars can be identified with Kun stars.  The rest of the Kun stars show obvious 
H$\alpha$ absorption or no emission, and 10 of them show the spectra of M-type stars in our grism spectra.   
Medium resolution spectroscopy by \cite{bal96} show that Kun 193 (No. 340; CepX-4)
exhibits conspicuous hydrogen emission, but 34 other Kun stars do not.  
They suggested that the lack of emission is due to the time variability of H$\alpha$ stars.  
\citet{con02} failed to detect H$\alpha$ emission in spectroscopic observations of
20 selected Kun stars, only finding emission in Kun 314 S. Thus, they suggest that the original 
identification of emission was incorrect. Our result is consistent with their suggestion.  

\citet{sic04,sic05}  extensively studied
the low-mass population of the two clusters in Cep OB2, Trumpler 37 and NGC 7160, to investigate
the disk evolution at ages 1--10 Myr.  
For Trumpler 37, they have performed optical photometry and multifiber spectroscopy 
in a square about 45$\arcmin$ per side, centered at  HD 206267.
Approximately 45\% of the low-mass population indicate NIR excesses, 
suggesting heated dust in the circumstellar disk. 
Finally, \citet{sic06b} examined the membership using  H$\alpha$ emission, Li $\lambda$6707 absorption lines, 
and radial velocities as membership criteria.
They identified accretion properties in 170 members in the central part of IC 1396.
53\% of the members of this list are also detected in our survey.
The  equivalent width of the H$\alpha$ emission-line of our candidates is consistent with that of
\citet{sic05}, except for only a few stars. 
We have detected $\sim$ 90\% and  $\la$ 30\%  of  Class-II and  Class-III
objects in their list as H$\alpha$ stars. 
This result is naturally explained by the higher detectability of weak emissions 
in their high dispersion slit spectroscopy. 

Using archival data of {\it Chandra} and {\it Spitzer}, 
\citet{mer09} have identified 25 X-ray sources as cluster members, which have 
MIR colors of CTTS or WTTS in the 10$\arcmin$ $\times$ 8$\arcmin$ region around HD 206267.  
Their sources consist of five CTTS and 20 WTTS.
Although four of the CTTS were detected in our observations,
only two WTTS match our H$\alpha$ stars. 

\citet{get07} obtained X-ray images of the bright rimmed cloud SFO 38 (IC 1396N) using the 
{\it Chandra} ACIS detector.  
They identified X-ray sources in the area of a square of 5$\arcmin$ with the {\it Spitzer} IRAC sources 
and concentrated their study on 25 sources associated with the globule SFO 38. 
Of the 117 X-ray sources, 17 sources are identified with our H$\alpha$ stars and five of these X-ray
sources are associated with H$\alpha$ stars in the SFO 38 cloud.

\subsection{Comparison with {\it IPHAS} Sources }

In the recent paper by \citet{bar11}, 
they identified 158 PMS candidates in a 7 deg$^2$ area towards IC 1396
by using the data of IPHAS \citep[the Isaac 
Newton Telescope/Wide Field Camera Photometric H$\alpha$ Survey; ][]{dre05}.
Out of 143 PMS candidates identified by \citet{bar11} in our field of view,
119 stars are in common with our H$\alpha$ sample.
We examined the common H$\alpha$ stars in IPHAS to confirm our photometric quality.   
The IPHAS magnitude is based on the Vega magnitude and hence we converted it to 
AB magnitude \citep{gon08}.  
\citet{bar11} restricted their sample to sources with 
$r'_{IPHAS}$ $<$ 20 mag, which corresponds to
$i'$  $<$ 19 mag if we adopt the colors of known T Tauri stars in their figure 4. 
For most of the stars in common, our $i'$-magnitudes are consistent with those in the IPHAS catalog within 0.1 mag.  
Although a large scatter ($\sigma$ = 0.3 mag) of the magnitude differences is found, it may be due to 
the photometric  variability of young stars.

\citet{bar11} compared their objects with the members confirmed by \citet{sic06b}, and estimate the completeness.
89\% of PMS candidates with the H$\alpha$ EW $>$ 30 {\AA} identified by \citet{sic06b} are 
recovered by \citet{bar11}.
We also estimated the completeness of our survey with the same manner.
For the PMS candidates including smaller EW ( $>$ 10 {\AA} ),  the fraction of recovered sources by \citet{bar11}
falls off to 50\%.
On the other hand, the fraction of recovered sources with EW $>$ 10 {\AA} maintains 86\%  (51 out of 59) in our sample.
We confirmed that none of their rejected candidates ( 30 stars ) by \citet{bar11} were included in our list.  
Thus, our observations are more sensitive than their survey for the emission-line stars with 
small H$\alpha$ equivalent width.

\subsection{Identification with {\it AKARI} Sources }


We identified 27 sources among our H$\alpha$ stars with the sources of  the point source catalog of 
the MIR All-Sky Survey obtained with the infrared camera (IRC) on board the {\it AKARI} satellite \citep{ish10},
with position differences of less than 5$\arcsec$.  
Table 3 gives the {\it AKARI}  9-$\micron$ ($S9W$) and 18-$\micron$  ($L18W$) magnitudes and their 2MASS photometric data.
Their positions on ($J$--$Ks$) vs. ($S9W$--$L18W$),  ($J$--$Ks$) vs. ($Ks$--$S9W$),  and ($J$--$Ks$) vs. ($Ks$--$L18W$)  diagrams 
are shown in Figure 3. The criteria for extracting TTS from these diagrams, proposed by  \citet{tak10},
are also shown as a dashed line (Figure 3a) or areas (Figure 3b,c).
The colors of three stars (Nos. 287, 309, 575) are consistent with cool carbon stars, 
and about 20 stars are pre-main sequence stars. 
The bright source No. 436 is the candidate for the exciting source of the HH 588 object.
Figure 4 shows the contour map of the number of emission-line stars overlaid on a false-color map 
using the IRC on  {\it AKARI}.
Blue is for 9 $\micron$ and red is for 18 $\micron$.
Probable members of 66 B--F stars from  \citet{con02}, supplemented by
8 OB stars from  {\it Hipparcos} results \citep{dez99}, 
are also shown as filled blue circles.
As for H$\alpha$ stars in  the West and East regions, high or intermediate-mass stars are concentrated
near the center of the large dust cavity, which is presumably swept up by the expansion of the H {\sc ii} region.
We note that their distribution is slightly shifted to the east of HD 206267.


\section{Analysis \& Discussion}

\subsection{Probable Members }

Our list of H$\alpha$ stars in Table 2 may contain foreground or background, magnetically active field stars.
We estimated the number of unrelated main 
sequence stars with enhanced chromospheric and coronal activity.  
Within the 4.2 deg$^2$ field towards the IC 1396, the Besancon model of the Galactic
stellar populations \citep{rob03} predicts $\sim$4000 and $\sim$ 1700 dM stars down to $V =$ 20 mag 
in the foreground and the background (the distance range between 870 and 1500 pc) field, respectively.   
Although 44\% of the foreground dM stars are in the range of $V <$ 19 mag, 3\% of the background
dM stars are in the same magnitude range.
Following the same approach as \citet{das05}, we used the 
the stellar luminosity function derived by {\it Hipparcos} \citep{jaw97}
and the dMe (with an H$\alpha$ EW $>$ 1 {\AA}) incidence from the nearby star survey of 
\cite{haw96}.  
We adopted a completeness limit for the present observations of $V =$ 20 mag. 
We assumed the scale height of 325 pc for an exponential falloff in the space density, and
0.7 mag kpc$^{-1}$ for the diffuse absorption.
The estimated number of dMe stars in the foreground and the background (up to 1500 pc)  is 193 and 33 
in our observed area, respectively.
The total number of  the field dMe stars, 54 per square degree, is consistent 
with the average number density of H$\alpha$ stars in the field area, 58 per square degree.

It is also possible that there is contamination by pre-main sequence stars associated with 
the foreground cloud \citep{pat95,wei96}.  
As the AKARI 9 $\micron$ image, which traces the photodissociation region in the periphery of the H {\sc ii} region (Figure 4), 
is likely to be dominated by polycyclic aromatic hydrocarbons (PAHs) 
emission features,
H$\alpha$ stars associated with 9-$\micron$ emission are not likely to be foreground objects.  
The dark clouds B 163 and B 163SW do not show any sign of bright rims and their radial velocities of 
the molecular emission line are 6 km s$^{-1}$ \citep{pat95}.  
The radial velocities of the other associated globules are between $-9$ km s$^{-1}$ and $+2$ km s$^{-1}$.  
From the radial velocity, \citet{leu82} suggested 150 pc for the distance of B 161, which  also shows no 
bright rims and no indication of heating from behind \citep{wei96}.   
The broad, curved dark lane Kh 161 \citep{kha60} and inverse S-shaped B 365, both of which show weak and diffuse emission 
at the northern and southern part of Figure 1, appear in foreground clouds.  


Figure 5 illustrates the NIR color--color diagram of our emission-line stars. 
The solid lines yield the locus of the main-sequence and of the giant branch 
according to \citet{bes88}, while the dash-dotted straight line indicates the locus 
of the unreddened CTTS following \citet{mey97}. 
We used the color transformation equations of \citet{car03} to convert their loci into the 2MASS  system. 
The reddening vector adopted from \citet{coh81} is also shown. 
We selected nearly 40\% of the H$\alpha$ stars whose NIR colors are consistent with the colors of 
the reddened CTTS in Figure 5 and calculated the source extinction by 
tracing back along the reddening line to the CTTS locus. 
In the above procedure, we have not separated between WTTS and CTTS candidates, 
and  treated all H$\alpha$ stars as the same way.
Although generally an unreddened WTTS does not show the color of the CTTS but rather of the normal main sequence,  
most WTTS have NIR colors of K- to early M-type stars, i.e.,  ($H$--$Ks$, $J$--$H$) $\sim$ (0.2, 0.64),
that is close to the locus of unreddened CTTSs.
Thus, we have not lost many WTTS in the above procedure.
Three carbon stars, discussed in the previous section and shown as open circles in Figure 5,  are excluded.
Figure 6 shows a histogram of the measured extinction of 252 stars excluding three carbon star candidates
(see \S 3.3) and No. 250/251.
The median value of the source extinction is $A$v$=$1.74 mag. 
This value is consistent with the average extinction $A$v$=$1.56 mag obtained by \citet{sic05} for Trumpler 37. 

The high mass stars in our surveyed area are two O-type stars, 
HD 206267 (O6) and HD 206183 (O9.5V),  and six early B stars (B0--B5) \citep{dez99}.  
When Rv$=$3.1, the foreground extinction of these OB stars should be  1.3--2.1 mag \citep{gar76}. 
\citet{con02} excluded stars with $A$v $<$ 1.0 mag as foreground stars and
excluded stars with  $A$v $>$ 2.3 mag as background objects.
As the background contamination by field dMe stars is not so high,
many of the emission-line stars  with $A$v $>$ 2.3 mag in our survey appear to be associated with
dark clouds or globules.
Therefore, we selected 189 stars with $A$v $>$ 1.0 mag as likely members associated with IC 1396. 
Most known T Tauri stars  \citep[e.g., ][]{sic06b}  in Table 2 have $A$v $>$  1.0 mag, but 12 T Tauri stars have low
extinction, and more than half of them are located near HD 206267. 


Figure 7 shows a histogram of the measured equivalent width.  
The fraction of weak (EW $<$ 10 {\AA}) emission-line stars among the IC 1396 PMS candidates
is low (27\%)  relative to that over the entire sample  (44\%).
It appears that the fraction of CTTS candidates among the PMS candidates in the IC 1396 region is higher than that among 
the candidates proposed as foreground (or background) objects.
We could not find any evidence of the difference between the spatial distribution of
CTTS and WTTS.

To identify the HAeBe stars, which have larger infrared excess than CTTS, 
we adopted the extended criteria in the NIR color--color diagram proposed by \citet{lel08}. 
They used the region between two parallel reddening lines on their figure 4 and above ($J$--$H$) $>$ 0.2 mag to 
pick up HAeBe stars, excluding  B[e] and classical Be stars 
from the young star sample as contaminants.  
According to their criteria, we selected five stars, redder than the reddening line passing through the truncated point of
the CTTS locus and $J$--$H$ $>$ 0.2 mag, including a known HAeBe, MVA-426 \citep{sic06b}.
It is suggested that from their  $J$ magnitudes  Nos. 13, 236, and 606 are HAeBe stars and
No. 479 and 545  are Class-I objects.
In the following discussion, we have added these stars and 12 T Tauri stars with low extinction to the probable members
totaling 205 stars.
The number of new unpublished PMS candidates identified in the current survey is 72
including three Class-I objects or HAeBe stars.


Figure 8 illustrates the contour map of the number density of emission-line stars associated with IC 1396.
Dashed boxes represent the eight subregions,  namely, the East, West, SFO38, SFO37, SFO39/41, Rim B, 
Halo, and Field regions.  
Their distribution is not very different from Figure 2, though
the peak of the distribution at the East region is slightly shifted westward.
While the aggregate at the north of B 365 (see Figure 1), which is shown in Figure 2b,  disappears in Figure 8
and therefore could be a foreground object, 
other aggregates should be associated with IC 1396. 
Table 4 summarizes the number of emission-line stars in each region.
The H$\alpha$ stars, whose $A$v values we could not obtain,  appear to be spread over the observed field rather uniformly.  
Thus, we suggest a large number of them are field stars.

\subsection{Ages and Masses of H$\alpha$ stars}

\citet{mar90} determined that the time after the first stars reach the main-sequence is 7 Myr in Trumpler 37 and 
\citet{sic05} confirmed 4 Myr for the age of Trumpler 37 
from optical photometry and theoretical isochrones. 
The expansion of the H {\sc ii} region creates a swept-up shell and results  in a compressed molecular ring around its periphery.  
From the size of the ring, \citet{pat95} suggested a dynamical age of the system of globules of 2--3 Myr.  


Figure 9 shows the extinction-corrected $i'$ vs. ($i'$--$J$) diagram of the 194 probable members
with $i'$-band data, using
the $i'$-band extinction correction from \citet{cad89}.
Here, we adopted the pre-main sequence model  of \citet{sie00}.
Their effective temperature and luminosity are converted into our system
using the table of \citet{ken95} and the equation of \citet{jor06}.
In Figure 9, the 0.1, 1, 3, 5, 10, and 100 Myr isochrones
are overlapped as well as  the evolutionary tracks for masses from 0.1 to 2.0 M$_{\sun}$, at an 
assumed distance of 870 pc. 
As the reddening vector runs almost parallel to the isochrones,
the extinction value would not severely affect our age estimate.

The uncertainty of the distance modulus (8.94--9.70)  and the choice of PMS models
affects the estimation of ages and masses of  H$\alpha$ stars from Figure 9.
We used the pre-main sequence models of \citet{pas99}, which are converted to our
photometric system by a table of \citet{ken95} and the equation of \citet{jor06}, for comparison. 
By assuming the nearest distance, 615 pc \citep{dez99}, the evolutionary tracks shift upwards, and it 
makes the age $\sim$0.2 dex older in both models.  
The age and mass histograms of the probable members in different choices of PMS models
adopting 870 pc are shown in Figure 10.
They indicate that the numbers of young stars ( $<$ 1 Myr ) predicted from the Siess models
(Fig 10a) are generally smaller than those predicted from the Palla \& Stahler models (Fig 10b).
On the other hand, the number of  low mass stars ( $<$ 0.4 M$_{\sun}$) derived from the Siess models 
(Fig 10c) are slightly large compared to those derived from the Palla \& Stahler models (Fig 10d). 
We should notice the incompleteness of our PMS candidate sample along with possible effects on the
distribution of masses and ages.
\citet{dar10}  carefully investigated the stellar population of the Orion Nebula Cluster by using these two 
PMS models.
The completeness function in the mass-age plane, derived from their statistical simulation for 
two evolutionary models, shows that a large incompleteness correction is needed in the low mass ($<$ 0.2 M$_{\sun}$) range. 
They  attributed the difference of results  to the different shapes of tracks and isochrones in this mass range.
Although we adopted the Siess models, the uncertainties would be larger for stars with ages  $<$ 1 Myr and masses $<$ 0.2 M$_{\sun}$.
Furthermore, we could not detect emission-line stars near the region of higher obscuration such as the globules,
and within 1$\arcmin$  of  the bright O star HD 206267, due to 
saturation of the detectors,  light contamination from this star, and diffraction spikes from this star.


Table 5 shows the number of H$\alpha$ stars in each age and mass bin in Figure 9. 
The age and mass histograms of H$\alpha$ stars in each subregion are shown in Figure 11.
We combined the SFO37/38/39/41/Rim B regions in the  BRC (bright-rimmed cloud) region.
Most of the H$\alpha$ stars have ages of $<$ 3 Myr and masses of 0.2--0.6 M$_{\sun}$. 
The number of stars with an age less than 1 Myr and with an age of 1--3 Myr are 
comparable in the region associated with bright rims, i.e., the BRC region.
We confirmed that the above result is also true for the age assigned from the Palla and Stahler model, 
and is not dependent on the choice of distance.
The histogram in the West region is similar to that of the BRC region,  probably because
most of the H$\alpha$ stars  in the West region are associated with Rim A.

From the extensive survey of young stars  by {\it Spitzer}, \citet{koe08} found two distinct
generations of star formation in W5.
They considered two triggered star formation mechanisms, radiative driven implosion \citep[RDI; e.g.,][]{mia09} and collect-and-collapse, 
to explain their findings and concluded that both mechanisms are at work in the W5 H {\sc ii} region.
\citet{deh10} extensively studied the nature of infrared bubbles which enclose H {\sc ii} regions and discussed 
star formation triggered by the expanding H {\sc ii} region at their edges.  
They show that most of these bubbles enclose H {\sc ii} regions ionized by O--B2 stars, many are formed to be
surrounded by cold dust shells (good candidates for C\&C processes),
and many have bright-rimmed dust condensations protruding inside the H {\sc ii} region
(good candidates for RDI process).
In the IC 1396 region, the presence of aggregates associated with the bright rims immediately inside the 
IC 1396 H {\sc ii} region or the near side of the ionizing 
star HD 206267, such as the West (Rim A), Rim B, and SFO38 regions, suggest that the star formation 
in these regions has advanced away from HD 206267 in $<$ 3 Myr.
Such a configuration is consistent with squeezing of the pre-existing condensations compressed 
by the pressure of the ionized gas.  
The kinematical structure of  SFO 37 \citep{sug97,ike08} and an X-ray study of SFO 38 \citep{get07} also favored
the  RDI model.

\subsection{Star Formation in the Central Region}

In our previous results  for the W5E H {\sc ii} region \citep{nak08}, 
we have found three aggregates of H$\alpha$ emission-line stars in the W5E H {\sc ii} region.
Two of them are located at the edge of the H {\sc ii} region, and one is near the central exciting O7 star. 
The former two aggregates are systematically younger than the latter. 

Based on a comparison of the stellar ages in the BRC region with those in the East region of IC 1396 (Figure 11),
it might appear  that the low-mass population in the East region is older.
We only find two stars younger than 1 Myr  in the East region, in contrast to the West region.
It may suggest that the star formation activity in the East region has ceased recently.
As there remains no molecular material associated with the population in the East region,  
the exciting star HD 206267 has already dissipated the natal molecular cloud. 
However, a Kolmogorov-Smirnov (K-S) test gives a probability of $>$ 15 \% 
that the two samples have the same distribution in age, even if we choose either PMS model. 
The difference between the ages of H$\alpha$ stars in the BRCs region and the East region is not 
statistically significant. 

\citet{sic06a} estimated ages for six members near Rim A or in the globule.
They found that the ages of five stars are younger than 1 Myr, which is significantly lower
than the average age of Trumpler 37. 
Figure 12 plots the spatial distribution of H$\alpha$ stars with derived ages.
At a glance, younger stars in Figure 12a are preferentially  located at the West region
compared to those at the East region.
Again we used K-S test to compare the distribution in right ascension and declination derived from
different choice of PMS models.  
A probability that  two samples of different age (Fig 12a and 12b)  have the same distribution in 
two coordinate axis is  $>$ 25 \% and $>$ 54 \%.
Thus, the age difference of young stars in the West and the East region is not statistically clear from the spatial distribution of our sample.

In addition to the West region, Figure 12b shows a number of stars with an intermediate age  (1--3 Myr) 
in the East and SFO38 regions, extending to  the Halo region. 
A linear sequence of young stars across the center is noticeable.
A filamentary dark cloud extends from Rim A to the western rim of the H {\sc ii} region. 
It is noteworthy that H$\alpha$ stars are distributed along this structure and further extend to the east of 
HD 206267.  This spatial distribution strongly suggests a 
fossil record of a star forming cloud including Rim A and B.
 
\citet{sic05}  argued a spatial east--west asymmetry in the young low-mass members of Trumpler 37, 
compared with more massive ones \citep{con02}. 
The same spatial asymmetry with respect to the central ionizing O star, HD 206267,  was seen 
in the distribution of the {\it Spitzer} IRAC excess objects \citep{sic06a}. 
The asymmetry of H$\alpha$ stars with high disk fraction may be explained by 
photo-evaporation of the disk material around the forming stars by strong UV radiation.
\citet{sic05} suggested that the interstellar material, lying between powerful O6 stars and 
young stars with disks, shields the disk from UV radiation.
However, there is no evidence of such an asymmetry of interstellar material from  molecular line \citep{pat95}
and millimeter dust continuum observations \citep{ros10}.
Actually, significant destruction of disks occurs for stars located out to distances 
of $<$ 0.5 pc, which corresponds to only a few arcmin from HD 206267, as found in NGC 2244 \citep{bal07}. 
\citet{sic05}  additionally suggested probable age differences for the explanation for the asymmetry in the
CTTS distribution. 
Instead of disk photo-evaporation,
we prefer that this spatial asymmetry represents the star formation sequence.
If most of the young low-mass population in the West region was formed as a result of a triggered event,
those in the East region could be formed along with massive stars.

\subsection{Star Formation in the BRC Region}

There are 116 probable members of emission-line star in the BRC region including the West region, and
the total mass  is about  60--70 M$_\sun$.
If we add 35 emission-line stars in the East region,  the total mass reaches 70-90  M$_\sun$.
Since  our survey is biased toward the region of low extinction,
the above estimates of the number and mass of the low-mass population in the BRC region including the West and the East region
should be considered the lower limits.
For example, \citet{cho10} found $\sim$45 YSOs in and around SFO 38, where we could not detect any 
H$\alpha$ stars in the globule.
Of these, 26 sources are Class-0/I, Class-I, and Class-I/II, but there are no such objects 
situated outside the ionized rim.  
It follows that the birth of low-mass stars associated with bright rims appears to be the primary mode of 
star formation in IC 1396 at present.

The higher mass O--F stars of Trumpler 37 \citep{dez99,con02}  
total $\sim$ 200 M$_\sun$ in our observed fields (Figure 4).
Assuming  the initial mass function of \citet{kro01}  between 0.01 M$_\sun$
and 50 M$_\sun$, we estimate the total mass of stars in Trumpler 37 to be $\sim$ 500 M$_\sun$.
Young low-mass stars, which probably formed by a triggered mechanism in the BRC region, 
contribute a fraction of the mass of stars in IC 1396.

\subsection{The Birth of HD 206267}

\citet{ike08} has determined the alignment of YSOs and their age gradient at the tip of SFO 37.
They argued that this alignment and the elongation of the head of the globule are
attributed to UV radiation from the O9.5V star, HD206183, at their formation stage.
The dominant exciting source, however, is currently HD 206267.
Consequently, HD 206267 formed  at the late stage of star formation
in the central part of the H {\sc ii} region and disrupted the natal molecular cloud.
\citet{bar11} reported the age gradient of the PMS candidates along the distance from this hot star.
Their results also favors the triggered formation of stars in IC 1396.  
HD 206267 is known as the Trapezium system of OB stars with a maximum separation of 15000 AU,
and the main component A is a triple \citep{sti95}.
\citet{bmm97} confirmed the observed masses and evolutionary model masses 
and suggested that the two bright components of the triple are O6.5 V((f))  (A$_1$) with 38--61 M$_\sun$ 
and O9.5: V (A$_2$) with  17--28 M$_\sun$.
Therefore, the age of HD 206267 is estimated to be $<$ 3 Myr from the main sequence lifetime \citep{sdk97}. 

Numerical simulation considering protostellar outflow-driven turbulence has been discussed 
by \citet{lin06} and \citet{nal07}.
Following their study, we suggested that the massive exciting star of the 
W5 E H {\sc ii} region was possibly formed after the continuous production of stars  \citep{nak08}.
The birth of the most massive star should be the last stage of star formation.
The  contiguous formation of stars for 10 Myr in Trumpler 37 could be supported by Figure 12.
Many of  the low-mass stars near  Rim A and SFO 38 might be formed in $\la$ 3 Myr.  
As there is no conspicuous molecular material except the elephant trunk globule, Rim A,  
in the evacuated dust cavity   (Figure 4),
HD 206267 was likely born in the central part after the continuous low mass star formation.
It is reasonable to suppose that the formation of massive stars and the associated low-mass  
population in Trumpler 37, continued for a few Myr.
Although it is difficult to separate the low-mass stars formed as the same generation with HD 206267
and formed by the trigger mechanism of HD 206267,
we believe that many of H $\alpha$ stars at the rim were triggered to form by a radiation driven implosion 
mechanism after the formation of HD 206267.

\section{Summary}

We have made an extensive survey of emission-line stars in the IC 1396 H {\sc ii} region to  investigate 
the low-mass PMS population  near the OB stars.
A total of 639 H$\alpha$ stars were detected over a 4-deg$^2$ area and their $i'$-photometry was performed.
Their spatial distribution shows several aggregates near the bright-rimmed clouds at 
the center and the edge of the H {\sc ii} region (Rim A/B and SFO 37, 38, 39, 41) and one near the main exciting star
of the H {\sc ii} region.
Based on the source extinction estimated from the NIR color--color diagram, we have selected PMS candidates
associated with IC 1396. The age and mass were derived from the extinction corrected color-magnitude diagram 
and theoretical evolutionary tracks.
Most of the H$\alpha$ stars have ages of $<$ 3 Myr and masses of 0.2--0.6 M$_\sun$. 

The spatial distribution of the PMS candidates indicates that the stellar population in the East region
is spatially distinct from that in the West region, associated with Rim A globule.
The age difference between two populations was also suggested by \citet{sic06b, bar11}.
The spatio-temporal gradients of the stellar age were also demonstrated near the bright-rimmed 
clouds by \citet{get07, ike08, cho10}. 
Our K-S test does not support the significant difference between ages of two populations.
Also, the spatial distribution of  $<$ 1 Myr stars is not significantly different from the distribution of 
1--3 Myr stars.
Although the star formation activity in the East region could be low in the last 1 Myr,
we need further confirmation.

The linear sequence of stars including the West (Rim A) region and Rim B region, which is conspicuous in 
Fig 12b  of 1--3 Myr age across the center of the H {\sc ii} region, suggests a fossil record of the star 
forming cloud. 
The formation of massive stars and associated low-mass population of Trumpler 37
continuing for a few Myr, and
the most massive star in Trumpler 37, HD 206267, was likely born $\la$ 3 Myr ago.   
Then, the IC 1396 H {\sc ii} region was formed by strong UV radiation.
In the last stage, the H {\sc ii} region presumably triggered the formation of low-mass stars in the dense 
inhomogeneity.
Our results may suggest that massive stars were born after the continuous formation
of  low-mass stars for 10 Myr.
The birth of the exciting star as the most massive one could be the late stage of 
slow but contiguous star formation in the natal molecular cloud.
It presumably triggered the formation of  many low-mass stars at the 
dense inhomogeneity in and around  the H {\sc ii} region  by a radiation-driven implosion.
Spectroscopic followup of our PMS candidates to establish their membership correctly 
and identify their nature  is needed to clarify the star formation history of  
this region.

\acknowledgments

The authors were supported by NAOJ for the use of the UH 2.2-m telescope for the observations.
This research made use of the NASA/IPAC Infrared Science Archive, operated by the 
Jet Propulsion Laboratory,
California Institute of Technology, under contact with the National Aeronautics and Space Administration.
This work was supported in part by a Grant-in-Aid for Scientific Research B (20403003) from the
Ministry of Education, Culture, Sports, Science and Technology.

\clearpage


\begin{figure*}
\epsscale{1.0}
\plotone{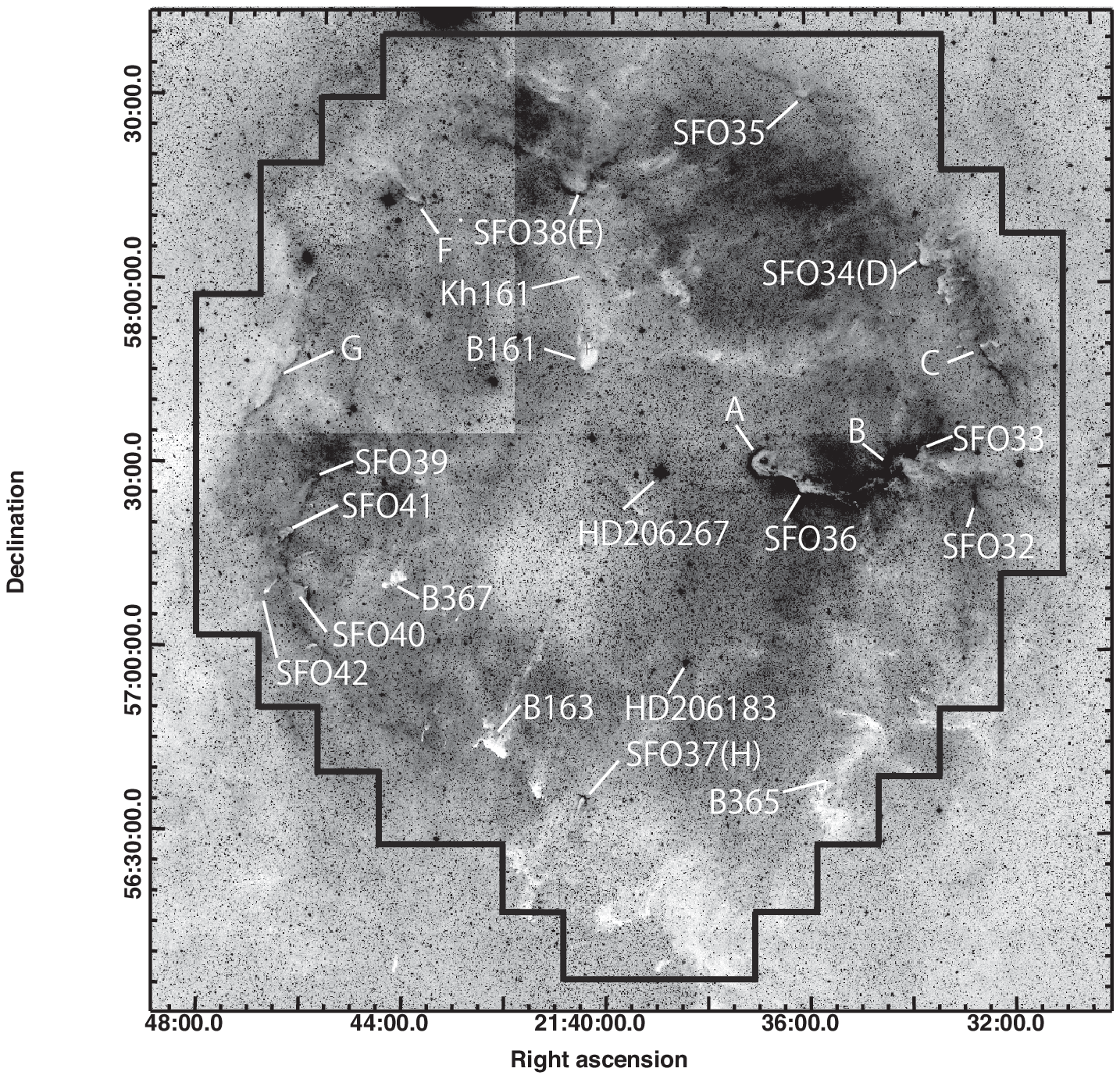}
\caption{Boundary of the observed fields shown on the Digitized Sky Survey R image of IC 1396. 
Two O stars,  bright-rims of IC 1396 (Rim A--H :\citealt{pot56}; SFO: \citealt{sfo91} ), 
and dark clouds (B:\citealt{b27}; Kh:\citealt{kha60}) are labeled by their name.
}
\end{figure*}

\begin{figure*}
\epsscale{1.0}
\plotone{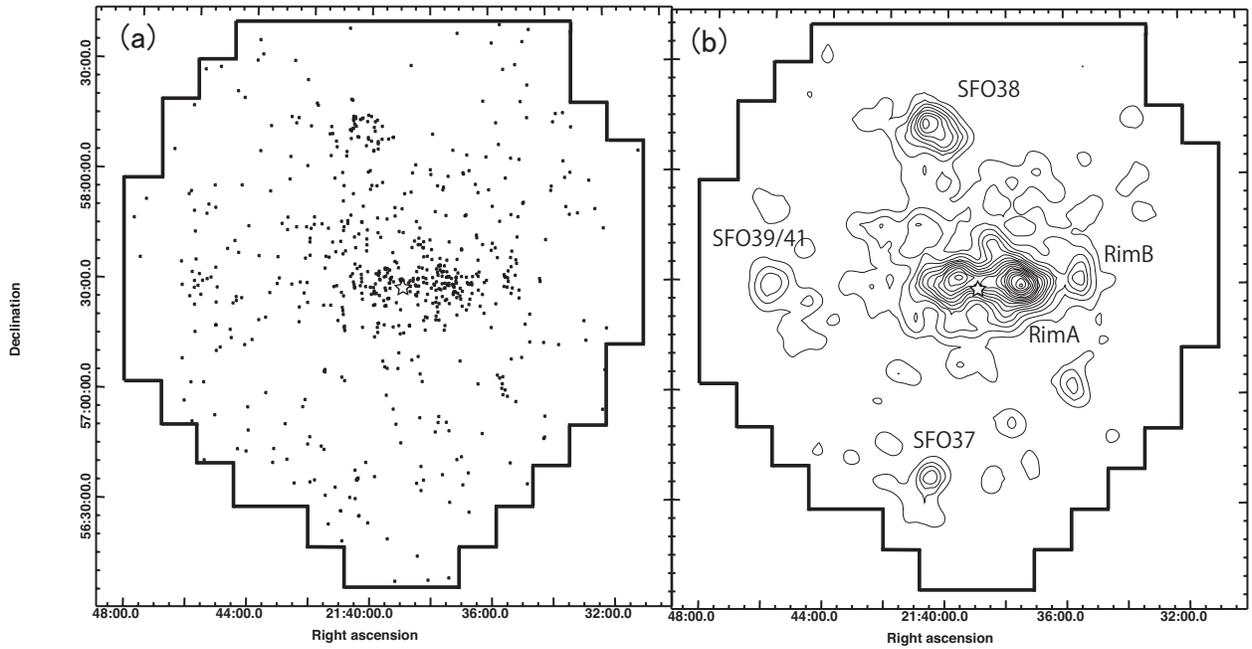}
\caption{(a) Spatial distribution of the detected emission-line stars in IC 1396.
(b) Contour map of the surface density of emission-line stars. 
The isolines are drawn at intervals of 150 stars per deg$^2$, starting from 100 stars per deg$^2$.  
The position of HD 206267 is shown by a star symbol  near the center, and the boundary of the 
observed fields is shown by a solid line.}
\end{figure*}

\begin{figure}
\epsscale{1.0}
\plotone{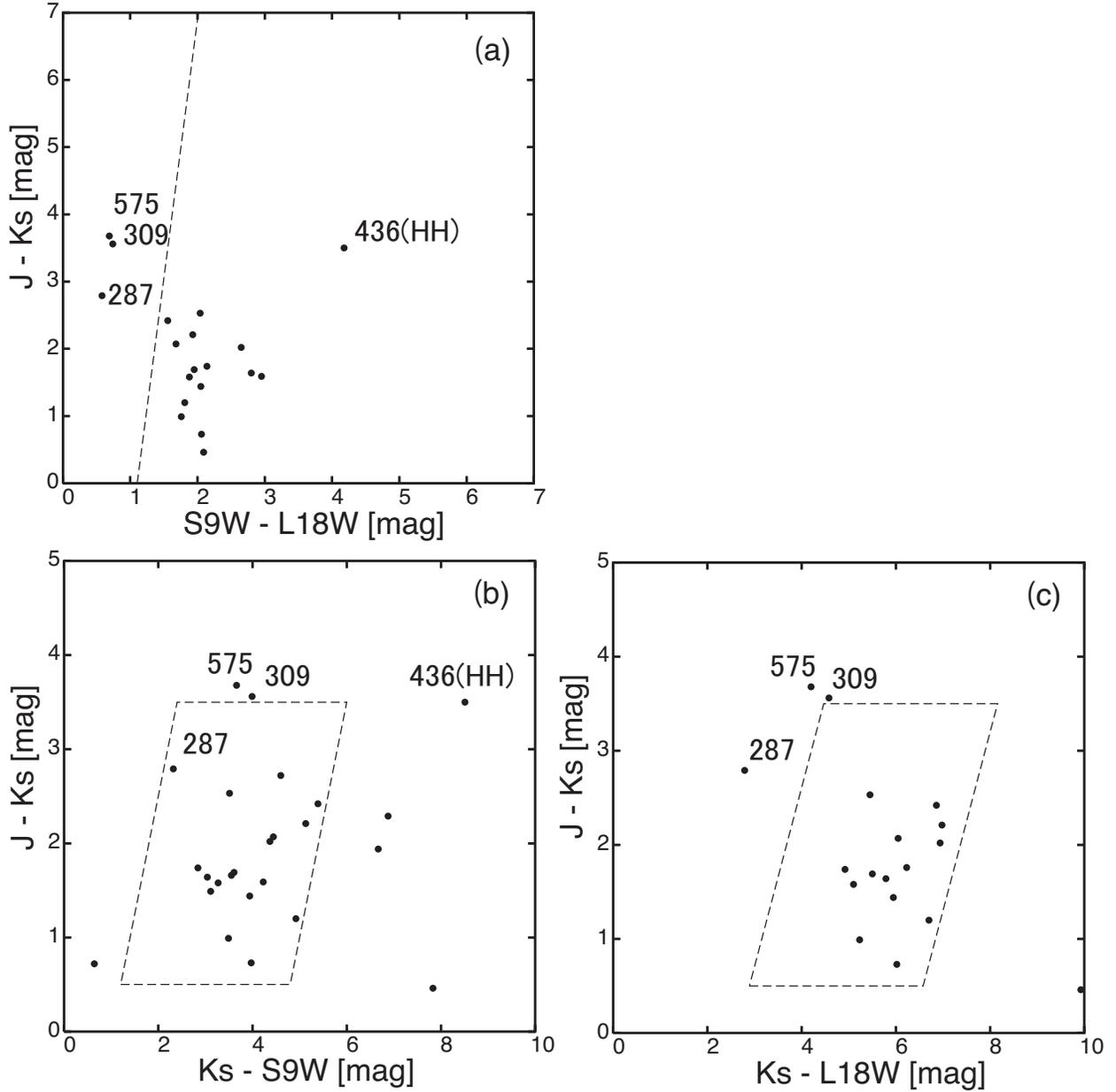}
\caption{(a) ($J$--$Ks$) vs. ($S9W$--$L18W$), (b) ($J$--$Ks$) vs. ($Ks$--$S9W$), and (c) ($J$--$Ks$) vs. ($Ks$--$L18W$) diagrams
of our H$\alpha$ stars identified in the {\it AKARI/IRC } mid-infrared point source catalog.
Three carbon stars and the candidate for the exciting source of the HH 588 object are labelled by their corresponding source numbers.
The dashed line in (a) is the criterion separating asymptotic giant branch stars from TTS adopted by \citet{tak10}.
The dashed areas in (b) and (c) indicate the areas occupied by TTS \citep{tak10}.   }
\end{figure}

\begin{figure*}
\epsscale{1.0}
\plotone{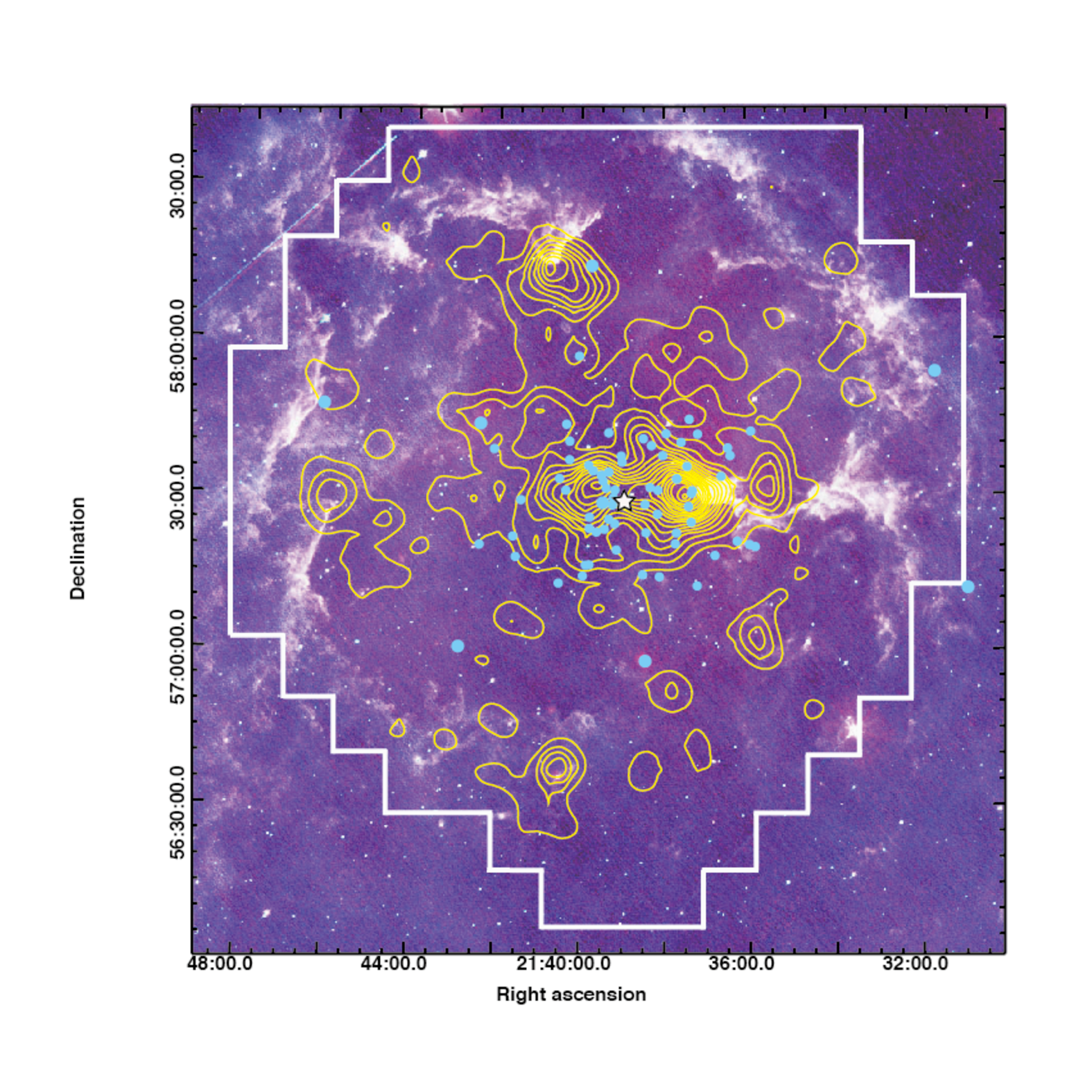}
\caption{Contour map of the number of emission-line stars overlaid on a false-color map 
using the infrared camera (IRC) on  {\it AKARI}.
Blue is 9 $\micron$ and red is 18 $\micron$.
O--F stars from  \citet{con02} and \citet{dez99} are also shown.
The star symbol near the center shows the position of the O star HD 206267.
The boundary of the observed fields is shown by the white solid line.}
\end{figure*}

\begin{figure*}
\epsscale{.80}
\plotone{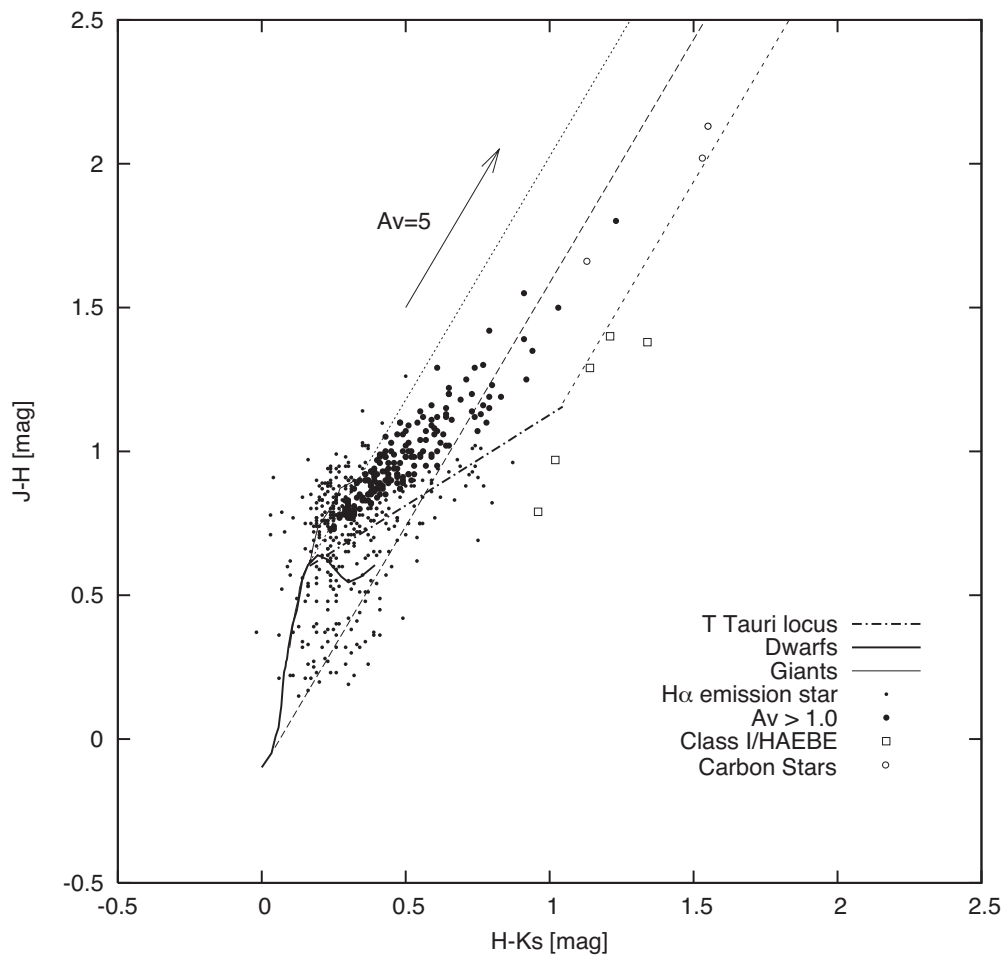}
\caption{NIR color--color diagram.
The thick and thin solid curves are the loci of dwarfs and giants, respectively. 
The data for the O9--M6 dwarf and G0--M7 giants are from \citet{bes88}. 
The dash-dotted line is the unreddened CTTS locus of \citet{mey97}. 
Emission-line stars are shown by dots, with the 
large dots representing emission-line stars with extinction $A$v $>$ 1.0.
Squares indicate Class-I or Herbig Ae/Be star candidates.
Three possible carbon stars are shown by open circles.
The $A$v$=$5 reddening vector is also shown.}
\end{figure*}

\begin{figure} 
\epsscale{.80}
\plotone{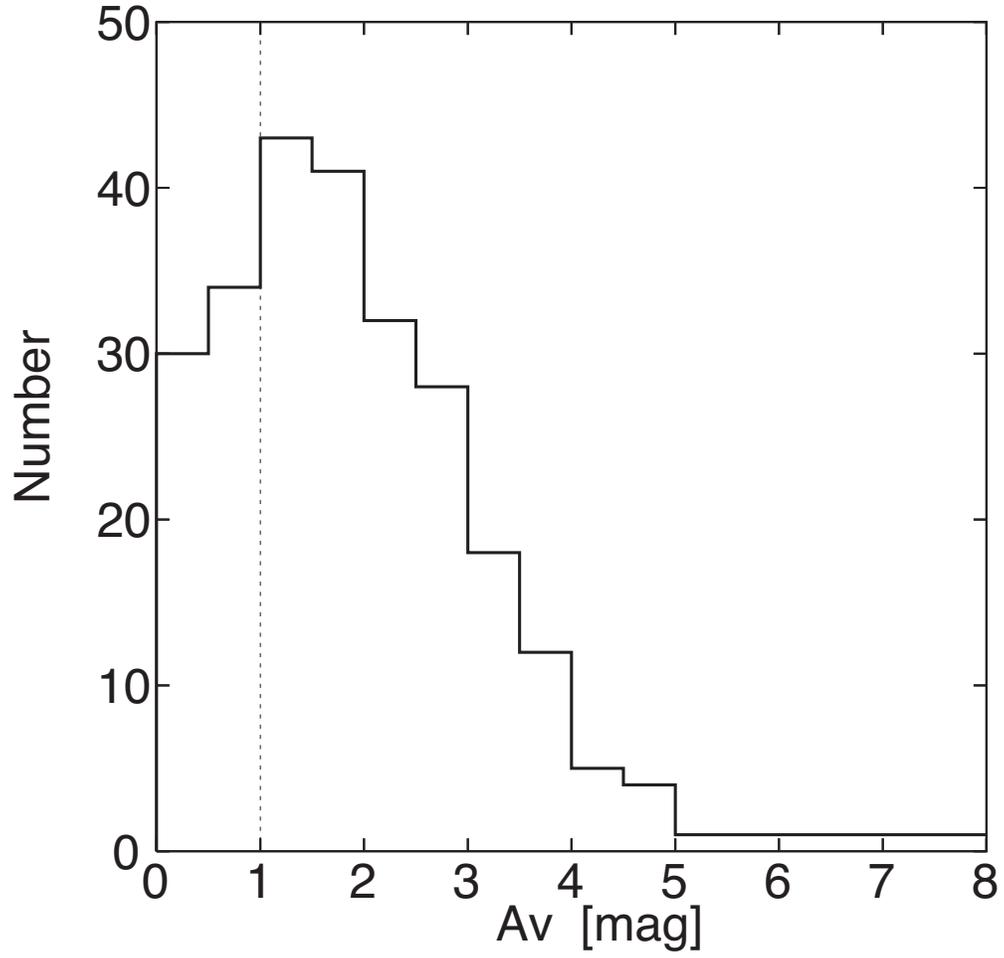}
\caption{Histogram of measured values of source extinction. 
The vertical line indicates $A$v $=$1.0.}
\end{figure}

\begin{figure} 
\epsscale{.80}
\plotone{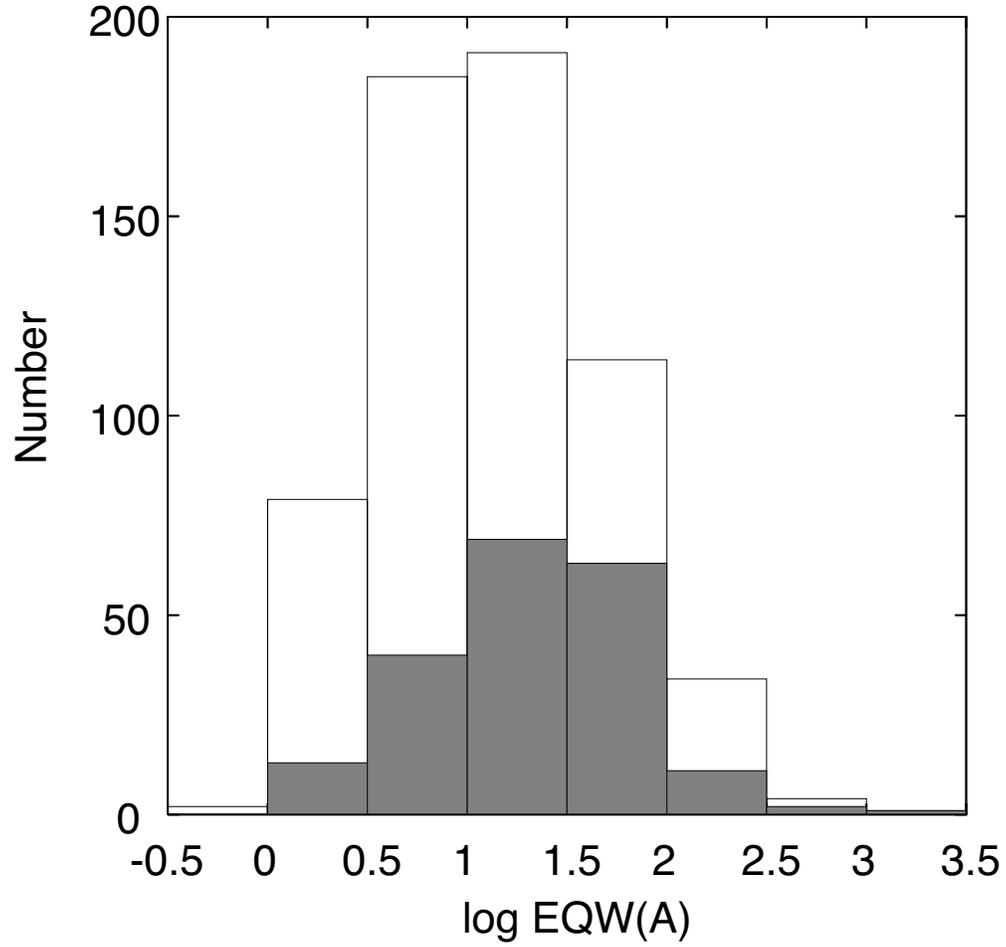}
\caption{Histogram of H$\alpha$ equivalent width.
The shaded inner histogram represents probable stellar members of the IC 1396 region.}
\end{figure}

\begin{figure*}
\epsscale{1.0}
\plotone{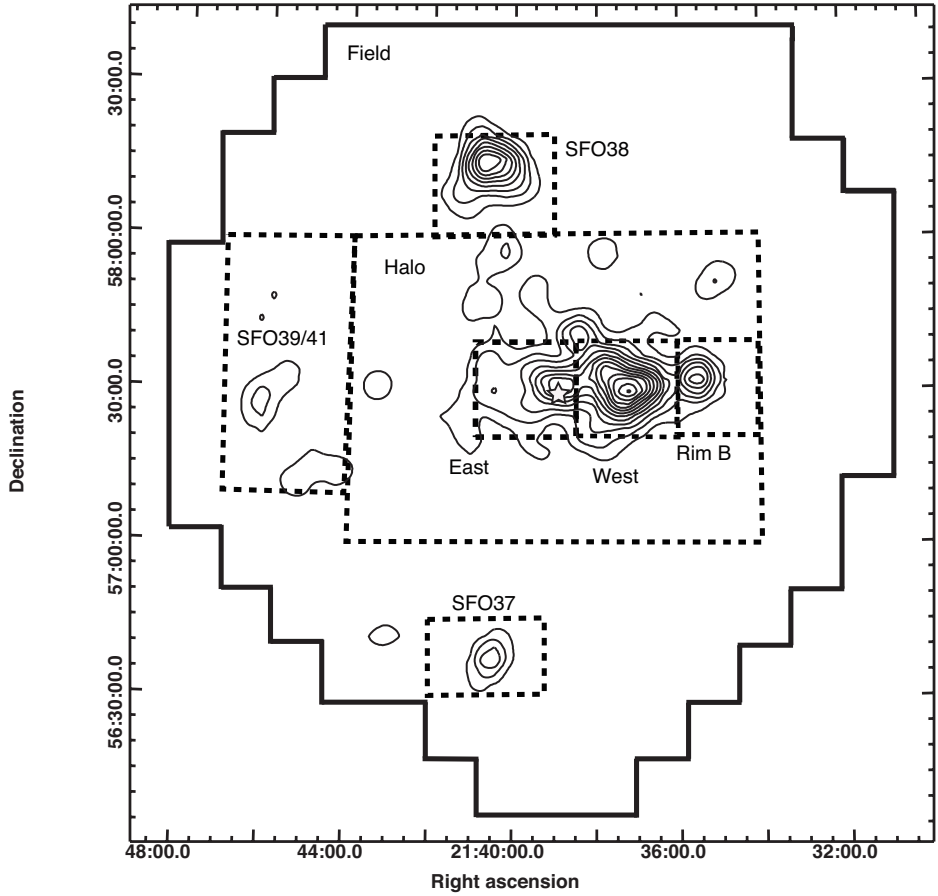}
\caption{Contour map of the number of probable members.
The isolines are drawn at intervals of 100 stars per deg$^2$, starting from 100 stars per deg$^2$.  
The dashed boxes represent the subregions referred to in the text.
The position of HD 206267 is shown by a star symbol near the center, and
the boundary of the observed fields is shown by a solid line.
}
\end{figure*}

\begin{figure*}
\epsscale{1.0}
\plotone{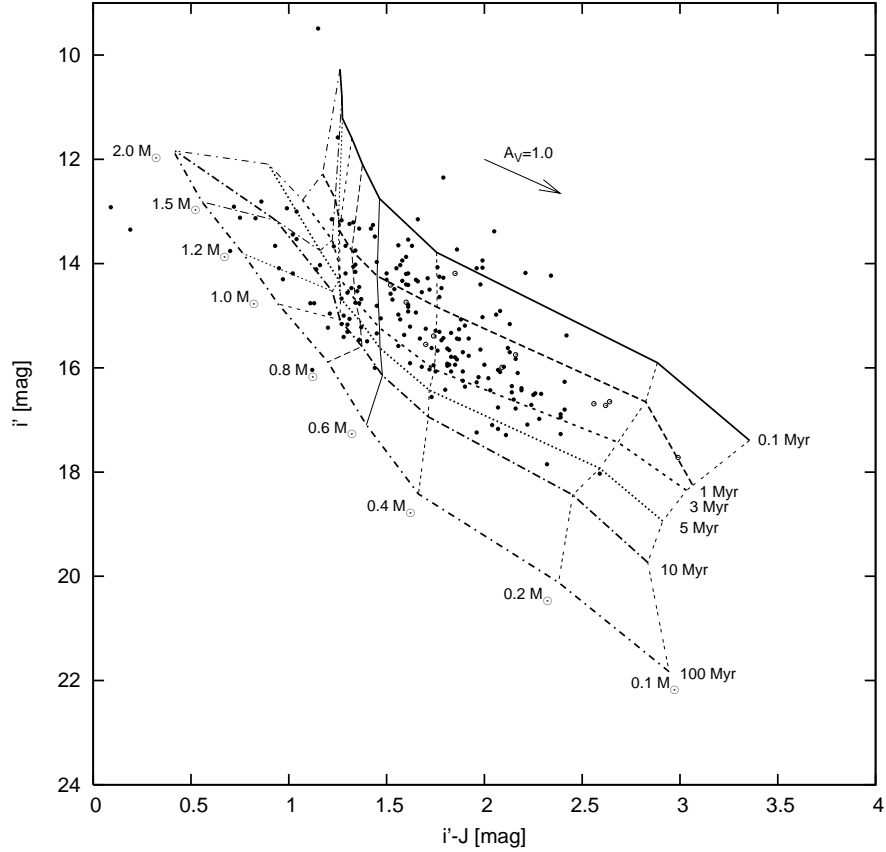}
\caption{Extinction-corrected $i'$ vs. ($i'-J$) diagram of probable members.
H$\alpha$ stars identified with known T Tauri stars with low extinction are shown by open circles.
The 0.1, 1, 3, 5, 10, and 100 Myr isochrones and the evolutionary
tracks for masses from 0.1 to 2.0 M$_\sun$ of \citet{sie00} are overlaid.
The reddening vector corresponding to $A$v$=$1.0 mag is also shown.}
\end{figure*}

\begin{figure*}
\epsscale{0.9}
\plotone{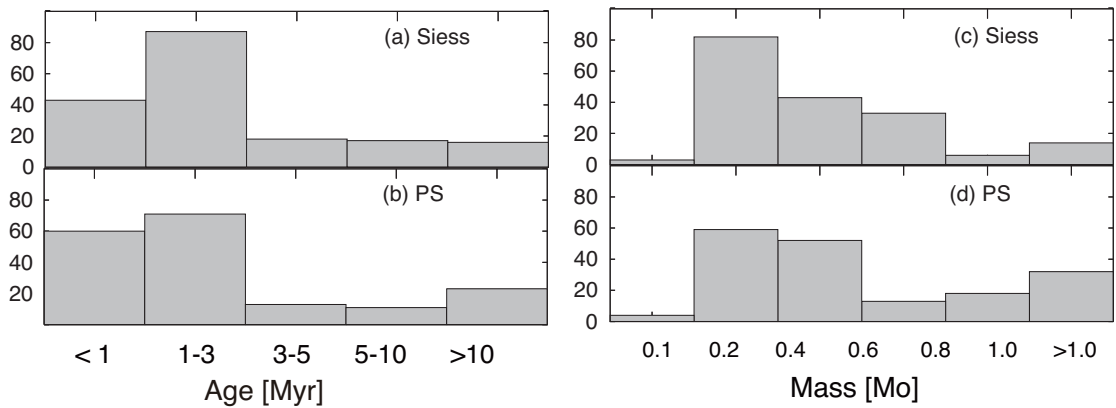}
\caption{Distribution of  ages (left panels) and masses (right panels) of H$\alpha$ stars predicted
from the choice of PMS models (\citet{sie00} and \citet{pas99}). 
}
\end{figure*}

\begin{figure} 
\epsscale{0.9}
\plotone{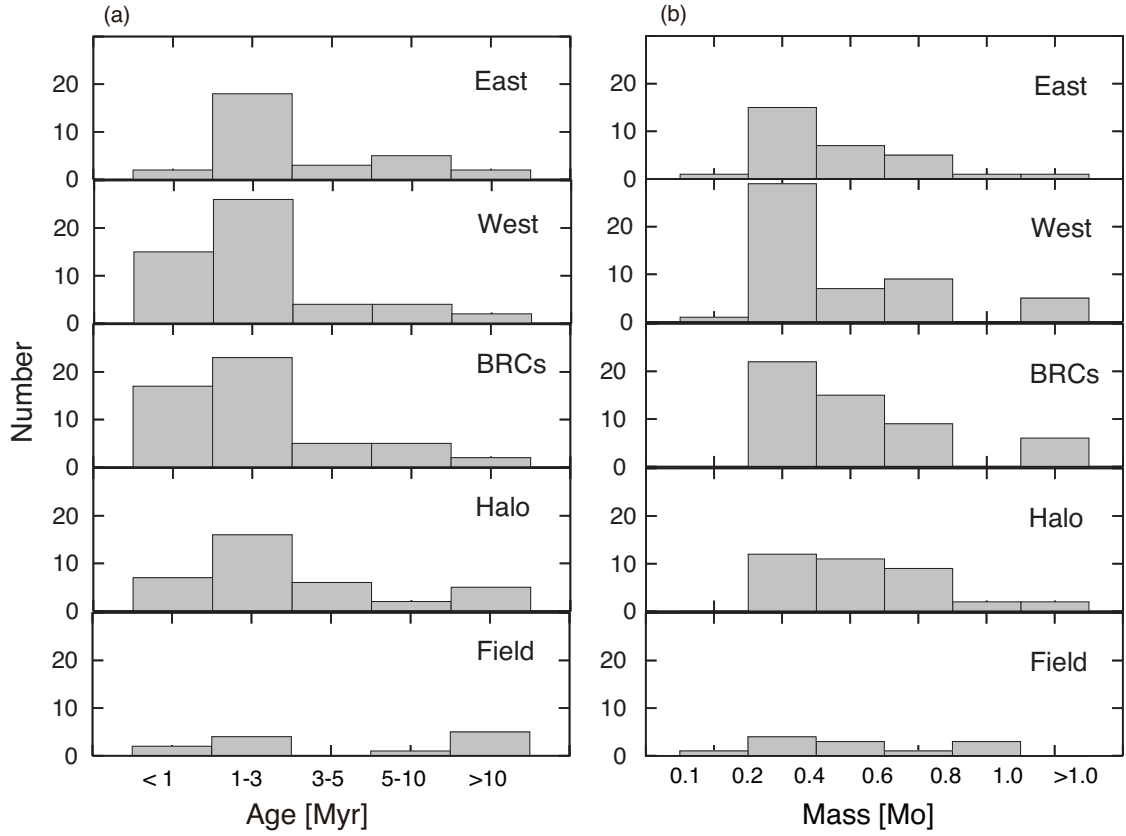}
\caption{Distribution of  (a) ages and (b) masses of H$\alpha$ stars in each of our
selected areas. The ages and masses are derived from a comparison with the Siess model in Figure 9.
The SFO37/38/39/41/Rim B regions are combined with the BRC region.}
\end{figure}

\begin{figure*}
\includegraphics[angle=90,scale=0.9]{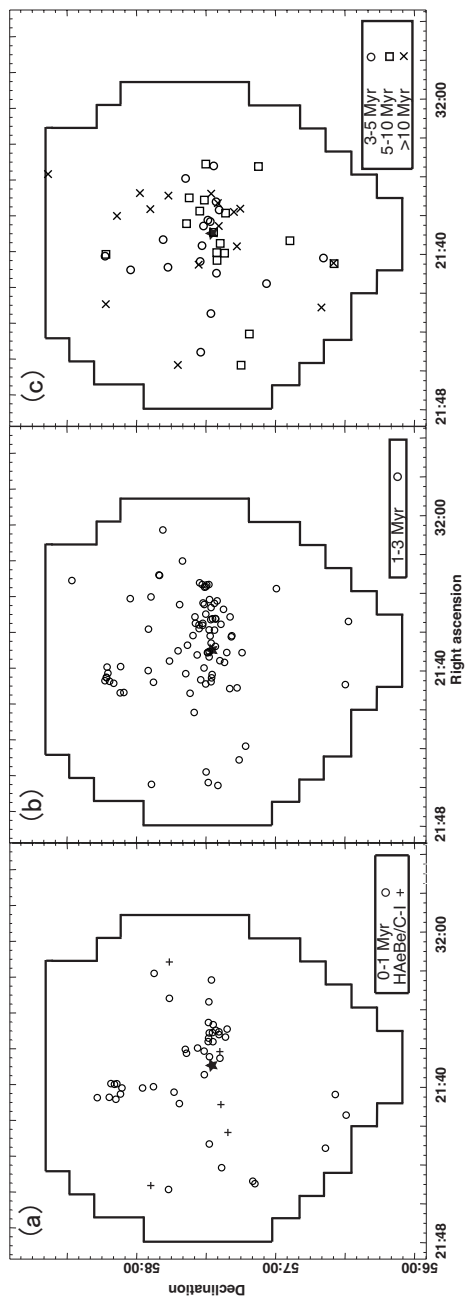}
\caption{Spatial distribution of H$\alpha$ stars of age (a) $<$ 1 Myr,
(b) 1--3 Myr, and (c) $>$ 3 Myr.  Class-I or Herbig Ae/Be stars selected from
Figure 5 are also shown by crosses in the left panel of (a). Stars with ages 3--5 Myr, 5--10 Myr, and
$>$ 10 Myr are shown by open circles, open squares, and crosses in the right panel of (c).
The star symbol near the center shows the position of HD 206267.}
\end{figure*}


\begin{table}
\begin{center}
\caption{Log of Observations.\label{tbl-1}}
\begin{tabular}{lrr}
\tableline\tableline
UT Date & number of fields & \\
                &    grism          & $i'$-band  \\
\tableline
2004 Nov 12                    & 1 &    \nodata  \\
2006 Sep 6                       & 2 &   \nodata  \\ 
2006 Nov 18, 20              & 5 &    \nodata \\	
2007 May 7                       & 2 &    \nodata \\
2008 Aug 1, 2, 4, 6, 7          & 52 &    54  \\
2008 Aug 8,9,11               & 38 &   \nodata \\
2008 Aug 21, 22, 24, 26  & 34 &  80 \\
2008 Nov 2                         & 2 &    \nodata \\
2009 Aug 13, 15, 17, 18   & 21 & 54 \\
\tableline
total                                      & 157& 188\tablenotemark{a} \\
\tableline
\end{tabular}
\tablenotetext{a}{including overlapped areas.}
\end{center}
\end{table}

\begin{deluxetable}{ccrrrrrrrrrl}
\rotate
\tablecaption{List of emission-line stars in IC1396 \label{tbl-1}}
\tablewidth{0pt}
\tablehead{
\colhead{No} & \colhead{RA} & \colhead{Dec} & \colhead{$i'$\tablenotemark{a}} & \colhead{EW[{\AA]}} &
\colhead{2MASS\tablenotemark{b}} & \colhead{quality flag}  & \colhead{$J$\tablenotemark{c}} & \colhead{$J$--$H$\tablenotemark{c}} &
\colhead{$H$--$Ks$\tablenotemark{c}} & \colhead{$A$v\tablenotemark{d}} &
\colhead{remarks\tablenotemark{e}} \\
\colhead{(1)} & \colhead{(2)} & \colhead{(3)} & \colhead{(4)} & \colhead{(5)} &
\colhead{(6)} & \colhead{(7)} & \colhead{(8)} & \colhead{(9)} & \colhead{(10)} &
\colhead{(11)} & \colhead{(12)}

}
\startdata
1 &  21:30:53.9 &  58:05:20.2 & 14.59 & 3 &  21305396+5805201 & AAA & 13.17 & 0.41 & 0.33 & \nodata &  \\
2 &  21:32:07.1 &  57:17:01.7 & 15.85 & 8 &  21320707+5717018 & AAA & 14.13 & 0.49 & 0.12 & \nodata &  \\
3 &  21:32:08.9 &  57:44:22.3 & -- & 5 &  21320901+5744224 & AAA & 14.42 & 0.72 & 0.34 & 0.14 &  \\
4 &  21:32:12.2 &  57:05:40.4 & 15.33 & 5 &  21321218+5705406 & AAA & 14.13 & 0.23 & 0.19 & \nodata &  \\
5 &  21:32:18.4 &  56:54:51.8 & 16.49 & 9 &  21321834+5654517 & AAA & 14.65 & 0.57 & 0.10 & \nodata &  \\
6 &  21:32:25.6 &  57:49:40.3 & 17.02 & 46 &  21322559+5749403 & AAA & 14.38 & 1.08 & 0.60 & 3.02 &  \\
7 &  21:32:36.9 &  57:57:27.8 & 15.30 & 3 &  21323682+5757276 & AAA & 13.84 & 0.54 & 0.16 & \nodata & \\
8 &  21:32:49.1 &  57:13:48.3 & 15.06 & 3 &  21324900+5713478 & AAA & 13.18 & 0.66 & 0.15 & \nodata &  \\
9 &  21:32:52.4 &  57:19:08.5 & 18.54 & 2 & \nodata &  \nodata & \nodata & \nodata & \nodata & \nodata &  \\
10 &  21:33:01.5 &  57:48:34.4 & 15.05 & 12 &  21330146+5748338 & AAA & 12.69 & 0.91 & 0.29 & \nodata &  \\
\enddata
\tablecomments{Table 2 is published in its entirety in the 
electronic edition of {\it Astronomical Journal}.  A portion is 
shown here for guidance regarding its form and content.}
\tablenotetext{a}{``s'' and ``--''  denote saturated and low photometric quality, respectively.}
\tablenotetext{b}{``6x'' represents 2MASS 6x Point Star Catalog}
\tablenotetext{c}{2MASS system. }
\tablenotetext{d}{calculated from the NIR color--color diagram}
\tablenotetext{e}{The  ``Class'' of the object,  Class-I (I), Class-II (II), Class-III (III) , 
transition object (TO), Herbig Ae/Be star (HAeBe), mainly from Sicilia-Aguilar et al. (2006b) 
are shown in the parenthesis. 
VES Coyne et al. (1983), [OH83] Ogura \& Hasegawa (1983), MVA Marschall \& van Altena (1987), 
[SBZ] Schulz et al. (1997), CGCS Alksnis et al. (2001), [OSP] Ogura et al. (2002), [RRY] Reach et al. (2004), 
[SHB] Sicilia-Aguilar et al. (2004), [SHH] Sicilia-Aguilar et al. (2005), [SHF] Sicilia-Aguilar et al. (2006b),
[SHW] Sicilia-Aguilar et al. (2007), [GFG] Getman et al. (2007), [ISW] Ikeda et al. (2008), [MMC] Mercer et al. (2009),
[BVD] the serial number of T Tauri candidates in Table B1 of Barentsen et al. (2011).}
\tablenotetext{f}{carbon star candidate}
\tablenotetext{g}{single 2MASS source}
\end{deluxetable}

\begin{deluxetable}{rrrrrl}
\tablecaption{AKARI and 2MASS photometric data of H$\alpha$ stars identified by AKARI PSC.\label{tbl-3}}
\tablewidth{0pt}
\tablehead{
\colhead{No} &  \colhead{$[S9W]$\tablenotemark{a}} & \colhead{$[L18W]$\tablenotemark{a}} & \colhead{$J$--$Ks$}  & \colhead{$Ks$}
& \colhead{remarks\tablenotemark{b}}

}
\startdata
13 & 4.44& 2.88 & 2.42 & 9.83 & LT Cep, IRAS 21317+5734 \\ 
21 & 5.72& 3.91 & 1.20 & 10.64 & \\ 
29 & 4.62& 2.69 & 2.21 & 9.75 &  \\ 
59 & 6.29& 4.24 & 1.44 & 10.22 & V488 Cep \\ 
60 & 5.81& \nodata & 1.94 & 12.49 & \\ 
81 & 6.99& \nodata & 1.66 & 10.54 & \\ 
115 & 5.88& 3.85 & 2.53 & 9.39 & [RRY] tet \\ 
121 & 6.58& 4.50 & 0.46 & 14.42 &  \\ 
127 & 6.77& 4.89 & 1.58 & 10.03 &  \\ 
203 & 5.87& 3.81 & 0.73 & 9.84 &  \\ 
236 & 5.93& 3.79 & 1.74 & 8.77 & MVA 426 \\ 
252 & 4.99& 3.04 & 1.69 & 8.59 & GM Cep \\ 
287 & 4.64& 4.06 & 2.79 & 6.96 & Kun 80, CGCS 5401\tablenotemark{c} \\ 
309 & 2.94& 2.20 & 3.56 & 6.94 &  \tablenotemark{c}\\ 
356 & 6.59& 4.83 & 0.99 & 10.08 &  \\ 
377 & 6.47& 4.73 & \nodata & 10.76 &  [SHF]93-361(II) \\ 
436 & 4.29& 0.11 & 3.50 & 12.80 & IRAS 21388+5622, HH 588 \\ 
456 & 3.80& \nodata & 2.29 & 10.67 & [OSP] BRC38 9 \\ 
487 & 7.43& \nodata & 0.72 & 8.07 & Kun 332 \\ 
575 & 1.17& 0.47 & 3.68 & 4.83 & CGCS 5434\tablenotemark{c} \\ 
596 & 6.80& 5.11 & 2.07 & 11.24 &  \\ 
606 & 7.07& \nodata & 2.72 & 11.67 &  \\  
611 & 6.61& 3.81 & 1.64 & 9.66 & IRAS 21440+5648\\ 
617 & 7.07& \nodata & 1.49 & 10.18 & Kun 215 \\ 
618 & \nodata & 5.03 & 1.76 & 11.32 & \\ 
621 & 6.26& 3.62 & 2.02 & 10.63 & \\  
622 & 6.63& 3.68 & 1.59 & 10.87 & \\  
\enddata
\tablenotetext{a}{The zero-magnitude flux densities for the AKARI bands are 56.262 Jy and 12.001 Jy for the $S9W$ and $L18W$ bands, respectively.}
\tablenotetext{b}{MVA Marschall \& van Altena (1987),  CGCS Alksnis et al. (2001), [OSP] Ogura et al. (2002), [RRY] Reach et al. (2004), 
[SHF] Sicilia-Aguilar et al. (2006b), [GFG] Getman et al. (2007)}
\tablenotetext{c}{carbon star candidate}
\end{deluxetable}

\begin{table}
\begin{center}
\caption{Summary of the numbers of H$\alpha$ stars in the surveyed area.\label{tbl-4}}
\begin{tabular}{llrr}
\tableline\tableline
  & sky area & number of H$\alpha$ stars\tablenotemark{a} & number density\\
  & deg$^{2}$    &                               & per deg$^{2}$ \\
\tableline
East                    & 0.11 &    89 (34) & 8.1  $\times$ 10$^{2}$ \\
West(Rim A)     & 0.11 & 116(56)  & 1.1 $\times$ 10$^{3}$\\ 
SFO38              & 0.13 &     51 (25) & 3.9 $\times$ 10$^{2}$ \\	
SFO37              & 0.10 &     15 (7)    & 1.6 $\times$ 10$^{2}$ \\
SFO39/41        & 0.34 &     43 (16) & 1.3 $\times$ 10$^{2}$ \\
Rim B                 & 0.09 &    23 (12) & 2.8 $\times$ 10$^{2}$ \\
Halo                  & 0.99 &  167 (38) & 1.7 $\times$ 10$^{2}$ \\
Field                 & 2.33 &  135 (17)  & 5.8 $\times$ 10$^{1}$ \\
\tableline
total                  & 4.2 & 639 (205) & \\
\tableline
\end{tabular}
\tablenotetext{a}{The number in parentheses  represents the probable members.
12 H$\alpha$ stars with low extinction, HAeBe/Class-I candidates (see \S 4.1), and
No. 436 (HH 588 center) are included as probable members.}
\end{center}
\end{table}

\begin{table}
\begin{center}
\caption{Ages and masses of H$\alpha$ stars associated with IC 1396. \label{tbl-5}}
\begin{tabular}{crrrrrr}
\tableline\tableline
Age/Mass   &  East & West & BRC & Halo & Field & total\\
\tableline
$<$ 1 Myr    & 2    & 15 &   17  & 7 & 2 &  43 \\
1--3 Myr        & 18 & 26  &  23  & 16 & 4 & 87 \\
3--5 Myr        & 3    &   4  &    5  &  6 &  0 &  18 \\
5--10 Myr      & 5    &   4 &    5  &  2 & 1 &   17 \\
10 Myr $<$  &   2 &  2  &     2  &  5 &  5 &    16 \\
\tableline \tableline
0.1--0.2 M$_\sun$  &   1  &   5  &   6   &  2 & 0 &  14  \\
0.2--0.4 M$_\sun$  &   1 &   0 &  0  &  2 & 3 & 6 \\
0.4--0.6 M$_\sun$  &   5 &  9 &  9  & 9 & 1 &  33 \\
0.6--0.8 M$_\sun$  &    7 &   7 &    15  &  11  & 3 &  43 \\
0.8--1.0  M$_\sun$ &    15 &   29 &     22  &  12  & 4 &  82 \\  
$>$ 1.0  M$_\sun$ &   1 &  1 &   0  &  0 &  1 &   3 \\
\tableline
total             & 30 & 51 & 52 & 36 & 12 & 181 \\
\tableline
\end{tabular}
\end{center}
\end{table}


\begin{thebibliography}{}
\bibitem[Allen et al.(2007)]{all07}
   Allen, L. et al. 2007, in Protostars and Planets V, ed. B. Reipurth, D. Jewitt, \& K. Keil (Tucson,
    Univ. Arizona Press), 361
\bibitem[Alksnis et al.(2001)]{alk01}
   Alksnis, A. et al. 2001, Baltic Astronomy, 10, 1
\bibitem[Bal\'{a}zs et al.(1996)]{bal96}
   Bal\'{a}zs, L.G. et al. 1996, \aap, 311, 145
\bibitem[Bessell  \& Brett (1988)]{bes88} Bessell, M., \& Brett, J.
    1988, \pasp, 100, 1134
\bibitem[Balog et al. (2007)]{bal07} 
    Balog, Z. et al.  2007, \apj, 660, 1532    
\bibitem[Barentsen et al. (2011)]{bar11} 
    Barentsen, G.  et al.  2011, \mnras, 415, 103    
\bibitem[Barnard (1927)]{b27} 
    Barnard, E.E. 1927,  A Photographic Atlas of Selected Regions of the Milky Way, 
    Carnegie Institution, Washington
\bibitem[Bonnet-Bidaud  \& Mouchet (1998)]{bon98} 
    Bonnet-Bidaud, J.M.  \& Mouchet, M. 1998, \aap, 332, L9   
\bibitem[Brice\~{n}o et al.(2007)]{bri07}
   Brice\~{n}o,~C. et al.
   2007, in Protostars and Planets V. ed.\ B. Reipurth, D. Jewitt, \& K. Keill 
   (Tucson: Univ. Arizona Press), p.345
\bibitem[Burkholder et al.(1997)]{bmm97}
   Burkholder, V., Massay, P.,  \& Morrell, N. 1997, \apj, 490, 328
\bibitem[Burningham et al.(2005)]{bur05}
   Burningham, B., Naylor, T., Littlefair, S.P., \& Jeffries,  R.D. \
   2005, \mnras, 363, 1389
\bibitem[Cardelli et al. (1989)]{cad89}
   Cardelli, J.A., Clayton, G.C., \& Mathis, J.S.  \
   1989, \apj, 345, 245
\bibitem[Choudhury et al. (2010)]{cho10}
  Choudhury, R., Mookerjea, B., \& Bhatt, H.C. 2010, \apj, 717, 1067
\bibitem[Cieza et al. (2005)]{cie05}
  Cieza, L.A., Kessker-Silacci, J.E., Jaffe, D.T., Harvey, P.M., \& Evans, N.J., II 
  2005, \apj, 635, 422.
\bibitem[Cohen et al. (1981)]{coh81}
  Cohen, J., Frogel, J.A., Persson, S.E., \& Elias, J.A. 1981, \apj, 249, 481
\bibitem[Contreras et al.(2002)]{con02}
  Contreras, M.E., Sicilia-Afguilar, A., Muzerolle, J., Calvet, N., Berlind, P., \& Hartmann, L. 2002, \aj, 124, 1585
\bibitem[Coyne et al.(1983)]{coi83}
  Coyne, G.V. \& MacConnell, D.J. 1983, Vatican Obs. Publ., 2, 73
\bibitem[Cutri et al. (2008)]{car03}
  Cutri, R.M.  et al.  2008, Explanatory Supplement to the 2MASS All Sky Data Release and Extended Mission Products,
   http://www.ipac.caltech.edu/2mass/releases/allsky/doc/sec6\_4b.html
\bibitem[Dahm \& Simon(2005)]{das05}
  Dahm, S.E. \& Simon, T. 2005, \aj, 129, 829
\bibitem[Da Rio et al.(2010)]{dar10}
  Da Rio, N. et al. 2010, \apj, 722, 1092
\bibitem[Deharveng et al. (2010)]{deh10}
  Deharveng, L.  et al. 2010, \aap, 523, 35 
\bibitem[de Zeeuw et al.(1999)]{dez99}
  de Zeeuw, P.T., Hoogerwerf, R., de Bruijne, J.H.J., Brown, A.G.A., \& Blaauw, A. 1999, \aj, 117, 354
\bibitem[Drew et al.(2005)]{dre05}
  Drew, J.E. et al. 2005, \mnras, 362, 753
\bibitem[Garrison \& Kormendy (1976)]{gar76}   
  Garrison, R.F. \& Kormendy, J. 1976, \pasp, 88, 865  
\bibitem[Getman et al. (2007)]{get07}   
  Getman, K.V., Feigelson, E.D., Garmire, G., Broos, P., and Wang, J. 2007, \apj 654, 316 [GFG]
\bibitem[Gomez et al. (1993)]{gom93}
   Gomez, M., Hartmann, L., Kenyon, S.J., Hewett, R.  \
   1993, \aj, 105, 1927
\bibitem[Gonz\'{a}lez-Solares et al. (2008)]{gon08}
   Gonz\'{a}lez-Solares, E.A. et al. 2008, \mnras, 388, 89
\bibitem[Hawley et al. (1996)]{haw96}
   Hawley, S., Gizis, J.E., \& Reid, I.N. 1996, AJ, 112, 2799
\bibitem[Hillenbrand  (1997)]{hil97}
   Hillenbrand, L.A. 1997, \aj, 113, 1733
\bibitem[Hughes \& Wouterloot  (1984)]{huw84}
   Hughes, V.A. \& Wouterloot, J.G.A. 1984, \apj, 276, 204
\bibitem[Ikeda et al.  (2008)]{ike08}
   Ikeda, H.  et al. 2008, \aj, 135, 2323 [ISW]
\bibitem[Ishihara et al.  (2010)]{ish10}
   Ishihara, D. et al. 2010, \aap, 514, A1
\bibitem[Jahrei{\ss} \& Wielen (1997)]{jaw97}
   Jahrei{\ss}, H.  \& Wielen, R. 1997, ESA-SP-402, 675
\bibitem[Jeffries et al. (2006)]{jef06}
   Jeffries, R.D., Maxted, P.F.L., Oliveira, J.M., \& Naylor, T.  2006, \mnras, 371, L6   
\bibitem[Jim\'{e}nez-Serra et al. (2007)]{Jim07}
   Jim\'{e}nez-Serra, I., Mart\'{i}n-Pintado, J., Rodr\'{i}guez-Franco, A., Chandler, C., Comito, C., Schilke, P.  2007, \apjl, 661, L187
\bibitem[Jordi et al. (1996)]{jor96}
   Jordi,C., Trullols, E.,  \& Galad\'{i}-Enr\'{i}quez, D.  1996, \aap, 312, 499
\bibitem[Jordi et al. (2006)]{jor06}
   Jordi, K., Grebel, E.K., \& Ammon, K. \
   2006, \aap, 460, 339
\bibitem[Kenyon \& Hartmann (1995)]{ken95}
   Kenyon, S.J. \& Hartmann, L. \
   1995,  \apjs, 101, 117 
\bibitem[Khavtassi (1960)]{kha60}
   Khavtassi, J.Sh. 1960, Atlas of Galactic Dark Nebulae, Abastumani Astrophys. Obs., Tbilisi
\bibitem[Koenig et al. (2008)]{koe08}
   Koenig, X.P. et al. 2008, \apj, 688, 1142
\bibitem[Kroupa (2001)]{kro01}
   Kroupa, P.  2001, \mnras, 322, 231   
\bibitem[Kun (1986)]{kun86}
   Kun, M. 1986, \apss, 125, 13
\bibitem[Kun (1987)]{kun87}
   Kun, M., Bal\'{a}zs, K.G., \& Toth, I. 1987, \apss, 134, 211   
\bibitem[Kun et al. (2008)]{kun08}
   Kun, M., Kiss, Z.T., \& Balog, Z. 2008, in Handbook of Star Forming Regions 1, 
   ed. B. Reipurth (Astronomical Society of  the Pacific), p.136
\bibitem[Kun \& P\'{a}sztor (1990)]{kup90}
   Kun, M. \& P\'{a}sztor, L. 1990, \apss, 174, 13 
\bibitem[Lada \& Lada (2003)]{lad03} 
   Lada, C.J.,  \& Lada, E.A. \
   2003,  \araa, 41, 57
\bibitem[Lee \& Lim (2008)]{lel08} 
   Lee, H.-T., \& Lim, J. 2008, \apj, 679, 1352 
\bibitem[Leung et al. (1982)]{leu82} 
   Leung, C.M., Kutner, M.L., \& Mead, K.N. 1982, \apj, 262, 583
\bibitem[Li \& Nakamura (2006)]{lin06} 
   Li, Z.,-Y.,  \& Nakamura, F. \
   2006,  \apj, 640, L187
\bibitem[Marschall \& van Altena (1987)]{mar87}
  Marschall, L.A., \& van Altena, W.F. 1987, \aj, 94, 71
\bibitem[Marschall et al. (1990)]{mar90}
  Marschall, L.A., Comins, N.F., \& Karshner, G.B. 1990, \aj, 99, 1538
\bibitem[Mercer et al. (2009)]{mer09}
  Mercer, E.P. et al.  2009, \aj, 138,7 [MMC]
\bibitem[Meyer et al. (1997)]{mey97}
  Meyer, M., Calvet, N., \& Hillenbrand, L.A. \
  1997, \aj,  114, 288
\bibitem[Miao et al. (2009)]{mia09}
  Miao, J., White, G.J., Thompson, M., \& Nelson, R.   2009, \apj, 692, 382 
\bibitem[Muench et al. (2008)]{mue08}
   Muench, A., Getman, K., Hillenbrand, L., \& Preibisch, T. 2008, in Handbook of Star Forming Regions 1,  
   ed. B. Reipurth (Astronomical Society of  the Pacific), p.483
\bibitem[Nakamura \& Li (2007)]{nal07}
  Nakamura, F.,  \& Li, Z.-Y. \
  2007, \apj,  662, 395  
\bibitem[Nakano et al. (1995)]{nak95}
  Nakano, M., Wiramihardja, S.D., \& Kogure, T. 1995, \pasj, 47, 889
\bibitem[Nakano et al. (2008)]{nak08}
  Nakano, M., Sugitani, K., Niwa, T., Itoh, Y., \& Watanabe, M. 2008, \pasj, 60, 739
\bibitem[Ogura \& Hasegawa (1983)]{ogu83}
   Ogura, K., \& Hasegawa, T. 1983, \pasj, 35, 299 
\bibitem[Ogura et al. (2002)]{ogu02}
   Ogura, K., Sugitani, K., \& Pickles, A. \
   2002, \aj, 123, 2597 [OSP]
\bibitem[Palla \& Stahler (1999)]{pas99}
   Palla, F., \& Stahler, S.W. \
   1999, \apj, 525, 772
\bibitem[Patel et al. (1995)]{pat95}
   Patel, N.A., Goldsmith, P.F., Snell, R.L., Hezel, T., \& Xie, T. 1995, \apj, 447,721
\bibitem[Pottasch(1956)]{pot56}
   Pottasch, S. 1956, Bull. Astron. Inst. Netherlands, 13, 77  
\bibitem[Pravdo et al. (2009)]{prv09}
   Pravdo, S.H., Tsuboi, Y., Uzawa, A., \& Ezoe, Y. 2009, \apj, 704, 1495
\bibitem[Reach et al. (2004)]{rea04}
   Reach, W.T. et al.  2004, \apjs, 154, 385 [RRY]
\bibitem[Robin et al. (2003)]{rob03}
   Robin, A.C., Reyle, C., Derriere, S., \& Picaud, S.  2003, \aap, 409, 523
\bibitem[Rosolowsky et al. (2010)]{ros10}
   Rosolowsky, E. et al.  2010, \apjs, 188, 123
\bibitem[Schaerer \& de Koter (1997)]{sdk97}
   Schaerer. D.,  \& de Koter, A. 1997, \aap, 322, 598
\bibitem[Schulz et al. (1997)]{sch97}
   Schulz,N.S., Bergh\"{o}fer, T.W., \& Zinnecker,H. 1997, \aap, 325, 1001
\bibitem[Sicilia-Aguilar et al. (2004)]{sic04}
   Sicilia-Aguilar, A., Hartmann, L., Brice\~{n}o, C., Muzerolle, J., \& Calvet, N. 2004, \aj, 128, 805
\bibitem[Sicilia-Aguilar et al. (2005)]{sic05}
   Sicilia-Aguilar, A., Hartmann, L., Hernandez, J., Brice\~{n}o, C., \& Calvet, N. 2005, \aj, 130, 188
\bibitem[Sicilia-Aguilar et al. (2006a)]{sic06a}
   Sicilia-Aguilar, A. et al.  2006a, \apj, 638, 897
\bibitem[Sicilia-Aguilar et al. (2006b)]{sic06b}
   Sicilia-Aguilar, A. et al.  2006b, \aj, 132, 2135 [SHF]
\bibitem[Sicilia-Aguilar et al. (2007)]{sic07}
   Sicilia-Aguilar, A. , Hartmann, L.W., Watson, D.,  \& Bohac, C.  2007, \apj, 659, 1637
\bibitem[Siess et al. (2000)]{sie00}
   Siess, L., Dufour, E., \& Forestini, M. 2000, \aap, 358, 593   
\bibitem[Smith et al. (2002)]{smi02}
   Smith, J.A. et al. \
   2002, \aj, 123, 2121
\bibitem[Stickland (1995)]{sti95}   
   Stickland, D.J. 1995, Observatory, 115, 180   
\bibitem[Sugitani et al. (1991)]{sfo91}
   Sugitani, K., Fukui, Y.,\&  Ogura, K. \
   1991, \apjs, 77, 59
\bibitem[Sugitani et al. (1997)]{sug97}
   Sugitani, K., Morita, K.I., Nakano, M., Tamura, M., \& Ogura, K. 1997, \apj, 486, L141
\bibitem[Takita et al. (2010)]{tak10}
   Takita, S. et al. 2010,  \aap, 519, A83
\bibitem[Tout et al. (1999)]{tou99}
   Tout, C.A., Livio, M., \& Bonnell, I.A. 1999, \mnras, 310, 360
\bibitem[Uehara et al. (2004)]{ueh04}
   Uehara, M. et al. \
   2004, Proc. SPIE, 5492, 661
\bibitem[Wang et al. (2010)]{wan10}
   Wang, P., Li, Z.-Y., Abel, T.,  \& Nakamura, F. 2010, \apj, 709, 27
\bibitem[Weikard et al. (1996)]{wei96}
   Weikard, H., Wouterloot, J.G.A., Castets, A., Winnewisser, G., \& Sugitani, K. 1996, \aap, 309, 581
\end{thebibliography}
\end{document}